\documentstyle[12pt]{article}

\def\emline#1#2#3#4#5#6{%
       \put(#1,#2){\special{em:moveto}}%
       \put(#4,#5){\special{em:lineto}}}
\def\newpic#1{}

\def\ti{\times}
\def\2{\frac{1}{2}}

\def\a{\alpha}
\def\b{\beta}

\def\m{\mu}
\def\n{\nu}

\def\ep{\epsilon}

\def\r{\rho}

\def\O{\Omega}

\def\N4{${\cal N }=4$}

\def\tr{\mathop{\rm tr}\nolimits}

\def\const{\mathop{\rm const}\nolimits}
\def\diag{\mathop{\rm diag}\nolimits}

\def\sgn{\mathop{\rm sgn}\nolimits}

\def\beq{\begin{equation}}
\def\eeq{\end{equation}}
\newcommand{\bea}{\begin{eqnarray}}
\newcommand{\eea}{\end{eqnarray}}
\def\ba{\beq\new\begin{array}{c}}
\def\bal{\begin{array}{l}}
\def\eal{\end{array}}
\def\bac{\begin{array}{c}}
\def\eac{\end{array}}
\def\ea{\end{array}\eeq\\}
\def\be{\ba}
\def\ee{\ea}

\makeatletter
\newdimen\normalarrayskip              
\newdimen\minarrayskip                 
\normalarrayskip\baselineskip
\minarrayskip\jot
\newif\ifold             \oldtrue            \def\new{\oldfalse}
\def\arraymode{\ifold\relax\else\displaystyle\fi} 
\def\eqnumphantom{\phantom{(\theequation)}}     
\def\@arrayskip{\ifold\baselineskip\z@\lineskip\z@
     \else
     \baselineskip\minarrayskip\lineskip2\minarrayskip\fi}
\def\@arrayclassz{\ifcase \@lastchclass \@acolampacol \or
\@ampacol \or \or \or \@addamp \or
   \@acolampacol \or \@firstampfalse \@acol \fi
\edef\@preamble{\@preamble
  \ifcase \@chnum
     \hfil$\relax\arraymode\@sharp$\hfil
     \or $\relax\arraymode\@sharp$\hfil
     \or \hfil$\relax\arraymode\@sharp$\fi}}
\def\@array[#1]#2{\setbox\@arstrutbox=\hbox{\vrule
     height\arraystretch \ht\strutbox
     depth\arraystretch \dp\strutbox
     width\z@}\@mkpream{#2}\edef\@preamble{\halign
\noexpand\@halignto
\bgroup \tabskip\z@ \@arstrut \@preamble \tabskip\z@ \cr}%
\let\@startpbox\@@startpbox \let\@endpbox\@@endpbox
  \if #1t\vtop \else \if#1b\vbox \else \vcenter \fi\fi
  \bgroup \let\par\relax
  \let\@sharp##\let\protect\relax
  \@arrayskip\@preamble}
\def\eqnarray{\stepcounter{equation}%
              \let\@currentlabel=\theequation
              \global\@eqnswtrue
              \global\@eqcnt\z@
              \tabskip\@centering
              \let\\=\@eqncr
              $$%
 \halign to \displaywidth\bgroup
    \eqnumphantom\@eqnsel\hskip\@centering
    $\displaystyle \tabskip\z@ {##}$%
    \global\@eqcnt\@ne \hskip 2\arraycolsep
         $\displaystyle\arraymode{##}$\hfil
    \global\@eqcnt\tw@ \hskip 2\arraycolsep
         $\displaystyle\tabskip\z@{##}$\hfil
         \tabskip\@centering
    &{##}\tabskip\z@\cr}
\begingroup\ifx\undefined\newsymbol \else\def\input#1 {\endgroup}\fi
\newfont{\hr}{msbm10}
\newfont{\ams}{msam10}

\font\numbers=cmss12
\font\upright=cmu10 scaled\magstep1
\def\stroke{\vrule height8pt width0.4pt depth-0.1pt}
\def\topfleck{\vrule height8pt width0.5pt depth-5.9pt}
\def\botfleck{\vrule height2pt width0.5pt depth0.1pt}
\def\Zmath{\vcenter{\hbox{\numbers\rlap{\rlap{Z}\kern 0.8pt\topfleck}\kern
2.2pt
                   \rlap Z\kern 6pt\botfleck\kern 1pt}}}
\def\Qmath{\vcenter{\hbox{\upright\rlap{\rlap{Q}\kern
                   3.8pt\stroke}\phantom{Q}}}}
\def\Nmath{\vcenter{\hbox{\upright\rlap{I}\kern 1.7pt N}}}
\def\Cmath{\vcenter{\hbox{\upright\rlap{\rlap{C}\kern
                   3.8pt\stroke}\phantom{C}}}}
\def\Rmath{\vcenter{\hbox{\upright\rlap{I}\kern 1.7pt R}}}
\def\Z{\ifmmode\Zmath\else$\Zmath$\fi}
\def\Q{\ifmmode\Qmath\else$\Qmath$\fi}
\def\C{\ifmmode\Cmath\else$\Cmath$\fi}
\def\R{\ifmmode\Rmath\else$\Rmath$\fi}
\def\numberbysection{\@addtoreset{equation}{section}
        \def\theequation{\thesection.\arabic{equation}}}
\numberbysection

\renewcommand{\theequation}{\thesection.\arabic{equation}}
\newcommand{\l@qq}[2]{\addvspace{2em}
 \hbox to\textwidth{\hspace{1em}\bf #1 \dotfill #2}}

\newcounter{app}

\def\app{\setcounter{equation}{0}
\def\theequation{\Alph{app}.\arabic{equation}}\par
   \addvspace{4ex}
   \@afterindentfalse
  \secdef\@app\@dapp}

\newcommand\@app{\@startsection {app}{1}{0ex}%
                                   {-3.5ex \@plus -1ex \@minus -.2ex}%
                                   {2.3ex \@plus.2ex}%
                                   {\normalfont\Large\bf}}
\def\@dapp#1{%
{\parindent \z@ \raggedright  \bf #1}\par\nobreak}
\def\l@app#1#2{\ifnum \c@tocdepth >\z@
    \addpenalty\@secpenalty
    \addvspace{1.0em \@plus\p@}%
    \setlength\@tempdima{8em}%
    \begingroup
      \parindent \z@ \rightskip \@pnumwidth
      \parfillskip -\@pnumwidth
      \leavevmode \bfseries
      \advance\leftskip\@tempdima
      \hskip -\leftskip
      #1\nobreak\hfil \nobreak\hb@xt@\@pnumwidth{\hss #2}\par
    \endgroup\fi}
\newcounter{sapp}[app]

\def\sapp{\def\theequation{\Alph{app}.\arabic{equation}}
\par
\@afterindentfalse
  \secdef\@sapp\@dsapp}
\newcommand{\@sapp}{\@startsection{sapp}{2}{\z@}%
                                     {-3.25ex\@plus -1ex \@minus -.2ex}%
                                     {1.5ex \@plus .2ex}%
                                     {\normalfont\large\bfseries}}

\def\@dsapp#1{%
{\parindent \z@ \raggedright  \bf #1
}\par\nobreak}
\newcommand{\l@sapp}{\@dottedtocline{2}{1.5em}{2.3em}}

\font\teneuf=eufm10  scaled  1440 
\font\seveneuf=eufm7 scaled  1\@ptsize00 
\font\fiveeuf=eufm5  scaled  1\@ptsize00 

\newfam\euffam
\textfont\euffam=\teneuf  \scriptfont\euffam=\seveneuf
  \scriptscriptfont\euffam=\fiveeuf

\def\hexnumber@#1{\ifnum#1<10 \number#1\else
 \ifnum#1=10 A\else\ifnum#1=11 B\else\ifnum#1=12 C\else
 \ifnum#1=13 D\else\ifnum#1=14 E\else\ifnum#1=15 F\fi\fi\fi\fi\fi\fi\fi}

\def\got{\ifmmode\let\next\got@\else
 \def\next{\errmessage{Use \string\got\space only in math mode}}\fi\next}
\def\got@#1{{\got@@{#1}}}
\def\got@@#1{\fam\euffam#1}

\newfont{\lgot}{eufm10 scaled 1440}%

\font\sevenmsa=msam6 
\newfam\msafam
\textfont\msafam=\sevenmsa
\def\hexnumber@#1{\ifnum#1<10 \number#1\else
\ifnum#1=10 A\else\ifnum#1=11 B\else\ifnum#1=12 C\else
\ifnum#1=13 D\else\ifnum#1=14 E\else\ifnum#1=15 F\fi\fi\fi\fi\fi\fi\fi}
\def\msa@{\hexnumber@\msafam}
\def\llcorner{\delimiter"4\msa@78\msa@78 }
\def\lrcorner{\delimiter"5\msa@79\msa@79 }
\mathchardef\blacktriangleright="3\msa@49
\mathchardef\blacktriangleleft="3\msa@4A
\mathchardef\trianglerighteq="3\msa@44
\mathchardef\trianglelefteq="3\msa@45
\font\tenmsb=msbm10 scaled 1\@ptsize00
\newfam\msbfam
\textfont\msbfam=\tenmsb
\def\msb@{\hexnumber@\msbfam}
\mathchardef\varkappa="0\msb@7B

\def\theequation{\thesection.\arabic{equation}}

\begin{document}

\begin{flushright}
ITEP-TH-46/01\\
LPTHE-01-55  \\
hep-th/0110178

\end{flushright}





\vspace{2.0cm}

\setcounter{footnote}{0}

\setcounter{equation}{0}

\centerline{
\huge Schwinger type processes }

\bigskip

\centerline{
\huge via branes and their gravity duals}
\vspace{1.0cm}
\centerline{ A. S. Gorsky$^{\,a,b}$, K. A. Saraikin$^{\,a,c}$ and K. G. Selivanov$^{\,a}$ }
\bigskip

\begin{center}
$^a$ {\em Institute of Theoretical and Experimental Physics, \\
B.Cheremushkinskaya 25, Moscow,  117259, Russia }\\
~\\
$^b$ {\em LPTHE, Universite' Paris VI, \\
4 Place Jussieu, Paris, France}\\
~\\
$^c$
{\em L.D.Landau Institute for Theoretical Physics, \\
Kosygina 2, Moscow, 117334, Russia }
\end{center}

\medskip \centerline{\rm e-mail: gorsky@heron.itep.ru,
saraikin@itp.ac.ru,  selivano@heron.itep.ru}

\bigskip

\begin{abstract}
We  consider Schwinger type
processes involving the creation of
the charge and monopole pairs
in the external
fields and
propose interpretation of
these processes  via corresponding brane
configurations in Type IIB string theory.
We suggest  simple description of some
new interesting nonperturbative processes like monopole/dyon transitions
in the electric field and W-boson decay in the magnetic
field using the brane language.
Nonperturbative pair production
in the strong coupling regime using the AdS/CFT correspondence
is studied.
The  treatment of the similar processes
in the noncommutative theories when
noncommutativity is traded for the
background fields is presented
and the possible role of the
critical magnetic field which is S-dual to the
critical electric field is discussed.

\end{abstract}


\bigskip

\bigskip

\section{Introduction}

Schwinger type processes are
the simplest nonperturbative phenomena
in the field theory. They include famous
charge/anticharge pair production in the constant
electric field \cite{schwin} as well as
the  monopole/antimonopole pair production in the magnetic
field  \cite{aff}. Natural
generalization of these processes in the string theory,
concerning the creation of
$p$-branes in $(p+2)$-form fields with the constant
curvature  has been developed in \cite{bt,Dowker}.
Another example
is the creation of the strings in the
constant electric field \cite{burgess,bp}
which reduces to the Schwinger pair production in the
field theory limit (see~\cite{amss} for a review).
One more class of the Schwinger type processes
includes so called induced processes, when
some external factors such as additional particles or branes,
finite temperature or chemical potential affect the
probability rate~\cite{CdL, AAM, VS, KR,GK, gs1}.
Very interesting phenomena
of the related nature has been recently discovered
by Myers \cite{myers}.

In this paper we  consider
Schwinger type processes from the point
of view of  theories on the brane worldvolumes.
It appears that the pair creation has simple
brane realization in Type IIB string
theory. For instance,
Euclidean configuration
corresponding to the electrically (magnetically) charged pair
creation in four dimensional U(2) gauge theory is described by the worldvolume of
F(D)-string  stretched between  two D3-branes.
Exponential dependence of the probability rate is given by the
minimum of the F(D)-string effective action.
In the simplest case of 1+1 dimensional U(2) gauge theory,
which is our basic example, this
effective action is just the area of the string
surface stretched
between two $(p,q)$-branes (see Fig.2).
This leads to the formula (\ref{rate2dim}) for the
probability rate.
When the external field value is small,
one gets well-known field theory result.
However, as the field becomes stronger,  the probability
gets modified in accordance with the
Bachas-Porrati result~\cite{bp}.
We
demonstrate that the production of the particles
in adjoint and in fundamental representation proceeds in a similar
manner.

The power of S-duality in type IIB string theory allows us
to treat Schwinger type processes in different electric-magnetic backgrounds on
the equal footing and suggests the existence of some
interesting processes on the field theory side. For instance,
we show that the monopole decays in the electric field (see Fig.1b, Fig7a)
while W-boson decays in the magnetic field (see Fig.7b). Evidently,
the probability rates of both processes are exponentially
suppressed but such  processes could have some cosmological applications.
Moreover, from this viewpoint it is natural to consider the
critical magnetic field (\ref{Bcrit}), which is S-dual to
the critical electric field: in the  critical magnetic field D-string
becomes effectively tensionless, while in the critical electric
field fundamental string  becomes effectively
tensionless.

One more problem addressed in this paper is the interpretation
of the Schwinger pair production via the gravity
duals of the large $N$ gauge theories in the spirit of \cite{mald}.
AdS/CFT correspondence gives one the
opportunity to take into account quantum effects in the gauge
theory by calculating the area of corresponding minimal surface in curved space.
We  consider two special brane configuration when the
bunch of $N$ D3-branes provides the $AdS_5$ geometry,
while one or two separate probe D3-branes are placed at finite
values of the radial coordinate. This corresponds to spontaneous
breaking of the gauge group down to $U(N) \times U(1)$ or $U(N) \times U(2)$.
Then we switch constant
electric field on the probe branes and calculate
the probability of the pair creation in the strong
coupling regime via the corresponding gravity solution.
In the case of the $U(N) \times U(1)$ gauge group we obtained the
formula (\ref{U1rate}) for the probability rate.
In the case of the $U(N) \times U(2)$ group we get results
in rather implicit form, since the situation turned out to be more
complicated there.
Note that unlike the Wilson loop calculation within
dual gravity picture \cite{malda, rey, esz,dg}
we have the massive parameter from the very
beginning and since we need this parameter to be finite, we do not
move probe D3-brane to infinity.
Another feature we found is
the absence of the Gross-Ooguri type phase transitions
\cite{ooguri,zarembo}
for the Schwinger type processes at the weak coupling
regime and the arbitrary field.

The theories on the D-brane worldvolume
with background electric and magnetic fields
in a certain limit amount to the NCYM theory \cite{sw} and
NCOS theory \cite{sst,gmms} theories
correspondingly. It is natural to
discuss the possible  role of the Schwinger
processes in these theories as the potential
sources of their nonperturbative instabilities.
We argue that such nonperturbative phenomena
are of some importance and especially
discuss the possible role of the critical
magnetic field.

The paper is organized as follows. Section 2 is devoted to calculation of
the pair production probability in the weak coupling regime via the
special brane configurations in Type IIB theory. Different
types of the electric-magnetic backgrounds are considered. The production of
the matter in the fundamental representation as well as the temperature
effects are discussed in short. In Section 3 we investigate some interesting processes
like decay of the monopole in the electric field and W-boson decay in the
magnetic field. These processes can be  considered both in the field
theory and in the brane setup. In Section 4 we use AdS/CFT correspondence to
calculate the amplitudes of the W-boson pair production in \N4 SUSYM theory
in the strong coupling limit. Section 5 is devoted to discussion on the
nonperturbative stability of NCYM and NCOS theories which are treated as the
theories on the D-brane worldvolumes with electric or magnetic backgrounds
correspondingly. Some remarks and speculations on the results obtained in
the paper and open questions can be found in Conclusion.

\section{Pair production via branes}

\subsection{Preliminaries}

Let us remind the key points concerning
the pair production in the  field theory. Consider
production of the electrically charged
pair in the electric field. Schwinger's
calculation~\cite{schwin} gives the following result
for the  probability of this process per unit time and volume
in the  four-dimensional U(1) gauge theory:

\be
\label{schw}
w=(gE)^2 \; \frac{2s+1}{8\pi^2}\sum_{n=1}^{\infty}
\frac{(-1)^{(2s+1)(n+1)}}{n^2}
\exp\left(-\frac{n\pi m^2}{gE}
\right)
\ee
where $s=0$ or $1/2$ is the particle spin, $g$ -- charge,
$m$ -- mass and $E$ -- constant
electric field. Later on we will mainly hunt for the
exponential factors. They can be derived by minimizing the
simple effective action which is natural in the first-quantized theory:

\be
S_{eff}= mL-gEA
\ee
where $L$ and $A$ are the perimeter and the area of the
closed particle trajectory in Euclidean space-time. Then

\be
w \propto \exp (-S_{eff}^{min})
\ee
From the symmetry considerations it follows that
this trajectory is just a circle of a certain radius $R$.
(see Fig.1a)
Therefore, the effective action reads

\be
S_{eff}=2\pi mR-\pi gER^2
\ee
and $R_{min}=m /(gE)$. Now we come to the answer

\be
w \propto \exp\left(-\frac{\pi m^2}{gE}\right)
\ee
for the probability rate, in agreement with the Schwinger formula (\ref{schw}).

\bigskip

\bigskip

\bigskip

\bigskip

\bigskip

\bigskip

\special{em:linewidth 0.4pt}
\unitlength 1.00mm
\linethickness{0.4pt}
\begin{picture}(130.33,73.33)
\put(28.67,29.00){\makebox(0,0)[cc]{(a)}}
\emline{74.67}{60.00}{1}{90.33}{60.00}{2}
\emline{114.67}{60.00}{3}{130.33}{60.00}{4}
\emline{90.67}{60.00}{5}{92.01}{57.94}{6}
\emline{92.01}{57.94}{7}{93.34}{56.12}{8}
\emline{93.34}{56.12}{9}{94.66}{54.54}{10}
\emline{94.66}{54.54}{11}{95.98}{53.19}{12}
\emline{95.98}{53.19}{13}{97.30}{52.08}{14}
\emline{97.30}{52.08}{15}{98.60}{51.21}{16}
\emline{98.60}{51.21}{17}{99.91}{50.57}{18}
\emline{99.91}{50.57}{19}{101.20}{50.17}{20}
\emline{101.20}{50.17}{21}{102.49}{50.00}{22}
\emline{102.49}{50.00}{23}{103.78}{50.08}{24}
\emline{103.78}{50.08}{25}{105.05}{50.38}{26}
\emline{105.05}{50.38}{27}{106.33}{50.93}{28}
\emline{106.33}{50.93}{29}{107.59}{51.71}{30}
\emline{107.59}{51.71}{31}{108.85}{52.72}{32}
\emline{108.85}{52.72}{33}{110.11}{53.97}{34}
\emline{110.11}{53.97}{35}{111.36}{55.46}{36}
\emline{111.36}{55.46}{37}{112.60}{57.19}{38}
\emline{112.60}{57.19}{39}{114.33}{60.00}{40}
\put(91.00,60.00){\circle*{1.33}}
\put(114.00,60.00){\circle*{1.33}}
\put(76.33,56.67){\makebox(0,0)[cc]{M}}
\put(102.33,29.00){\makebox(0,0)[cc]{(b)}}
\put(102.00,47.33){\makebox(0,0)[cc]{$e^-$}}
\put(127.67,57.00){\makebox(0,0)[cc]{M}}
\emline{28.67}{68.65}{41}{30.57}{68.43}{42}
\emline{30.57}{68.43}{43}{32.38}{67.81}{44}
\emline{32.38}{67.81}{45}{34.01}{66.80}{46}
\emline{34.01}{66.80}{47}{35.37}{65.46}{48}
\emline{35.37}{65.46}{49}{36.41}{63.85}{50}
\emline{36.41}{63.85}{51}{37.07}{62.05}{52}
\emline{37.07}{62.05}{53}{37.31}{60.15}{54}
\emline{37.31}{60.15}{55}{37.14}{58.25}{56}
\emline{37.14}{58.25}{57}{36.54}{56.43}{58}
\emline{36.54}{56.43}{59}{35.56}{54.78}{60}
\emline{35.56}{54.78}{61}{34.25}{53.40}{62}
\emline{34.25}{53.40}{63}{32.66}{52.33}{64}
\emline{32.66}{52.33}{65}{30.87}{51.64}{66}
\emline{30.87}{51.64}{67}{28.98}{51.36}{68}
\emline{28.98}{51.36}{69}{27.07}{51.50}{70}
\emline{27.07}{51.50}{71}{25.24}{52.06}{72}
\emline{25.24}{52.06}{73}{23.58}{53.01}{74}
\emline{23.58}{53.01}{75}{22.17}{54.30}{76}
\emline{22.17}{54.30}{77}{21.07}{55.88}{78}
\emline{21.07}{55.88}{79}{20.35}{57.65}{80}
\emline{20.35}{57.65}{81}{20.04}{59.54}{82}
\emline{20.04}{59.54}{83}{20.15}{61.45}{84}
\emline{20.15}{61.45}{85}{20.67}{63.29}{86}
\emline{20.67}{63.29}{87}{21.59}{64.97}{88}
\emline{21.59}{64.97}{89}{22.86}{66.40}{90}
\emline{22.86}{66.40}{91}{24.41}{67.52}{92}
\emline{24.41}{67.52}{93}{26.17}{68.28}{94}
\emline{26.17}{68.28}{95}{28.67}{68.65}{96}
\put(20.00,60.00){\circle*{1.49}}
\put(37.33,60.00){\circle*{1.33}}
\put(25.00,70.67){\makebox(0,0)[cc]{$e^+$}}
\put(24.67,49.00){\makebox(0,0)[cc]{$e^-$}}
\put(68.00,32.33){\vector(1,0){0.2}}
\emline{47.00}{32.33}{97}{68.00}{32.33}{98}
\put(57.33,70.67){\makebox(0,0)[cc]{E}}
\put(57.33,35.50){\makebox(0,0)[cc]{time}}
\put(54.00,73.33){\vector(0,1){0.2}}
\emline{54.00}{57.00}{99}{54.00}{73.33}{100}
\emline{91.00}{60.00}{101}{93.47}{60.39}{102}
\emline{93.47}{60.39}{103}{95.94}{60.68}{104}
\emline{95.94}{60.68}{105}{98.43}{60.88}{106}
\emline{98.43}{60.88}{107}{100.92}{60.98}{108}
\emline{100.92}{60.98}{109}{103.42}{60.99}{110}
\emline{103.42}{60.99}{111}{105.92}{60.91}{112}
\emline{105.92}{60.91}{113}{108.44}{60.73}{114}
\emline{108.44}{60.73}{115}{110.96}{60.45}{116}
\emline{110.96}{60.45}{117}{114.00}{60.00}{118}
\put(101.67,63.67){\makebox(0,0)[cc]{D}}
\end{picture}

\vspace{-1.7cm}

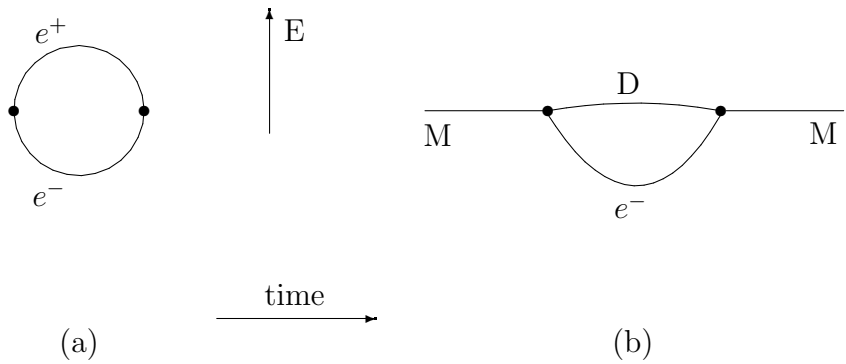
Fig.1: Field theory effective picture of the Schwinger type processes
in Euclidian space-time:
(a)  $e^+ e^-$ pair creation in constant electric field  $E$.
(b) Monopole decay into dyon and anti-charge in constant electric field:
$M \to D + e^-$. Since at weak coupling the dyon is much heavier then
$e^-$, its trajectory is almost the straight line.

\vspace{1.7cm}

In string theory there is a similar process: string pair production
in a constant electric field. Generalization of the Schwinger formula
(\ref{schw}) for this case  was obtained by
Bachas and Porrati~\cite{bp}. For the fermionic strings it has the form:

\newpage

\be
w=\frac{g E}{(2\pi)^{D-1}\ep}
\sum_{S}\sum_{n=1}^{\infty} (-1)^{(n+1)(a_S+1)}
\left(\frac{|\ep|}{n}\right)^{D/2}
\exp\left(-\frac{\pi m_S^2}{|\ep|}n\right)
\ee
Here $D$ is a number of non-compact dimensions,
$m_S$ is a mass of the  string state $S$ and\footnote{it is assumed that
$\a'=1/2$ in this formula}

\be
\ep=\frac{2}{\pi} {\rm arcth}(\pi gE)=
\frac{1}{\pi} \log \frac{1+\pi gE}{1-\pi gE}
\ee
Below we   describe how the exponential dependence of this probability
can be obtained from the effective string action $S_{eff}$.

\subsection{Stringy W-boson pair production in U(2) gauge theory}

Let us turn to the brane picture behind this
process and start with 1+1~dimensional example and U(2)
gauge group. The theory without external field is
realized on the worldvolume of two D1-branes,
displaced at the distance $v$ in the transverse direction
which corresponds to the vacuum expectation value
of the scalar. To switch on the electric field
one can add standard boundary term~\cite{ACNY}:

\be
\label{Sbndry}
\oint d\sigma E X^0 \partial_{\sigma} X^1
\ee
to the Nambu-Goto action, but in type IIB theory we can
consider dyonic $(n,1)$-strings
instead of the D1-branes. Then the value of the electric field in the
theory reads as (see, for instance, \cite{pole})

\be
\label{E}
E= \frac{ng_{st}}{\sqrt{1+ (ng_{st})^2}}
\ee
where $g_{st}$ is a string coupling constant, which is related to the
field coupling constant $g$ as $g_{st}=g^2$.
It follows from this formulae that there exists the critical
value of the electric field $E_{cr}=1$.
When the field reaches this value the effective tension of the
open string becomes zero.
From the field theory viewpoint this corresponds to the
pair production without exponential suppression
since
$S^{min}_{eff}[E_{cr}]=0$.
The W-bosons are represented by the F-strings stretched
between two dyonic strings. We assume that with respect
to the U(2) gauge group the electric field
is chosen as $\diag(E,-E)$.

\vspace{1cm}

\special{em:linewidth 0.4pt}
\unitlength 1.00mm
\linethickness{0.4pt}
\begin{picture}(155.67,109.67)
\emline{48.67}{99.67}{1}{54.51}{99.99}{2}
\emline{54.51}{99.99}{3}{59.34}{100.33}{4}
\emline{59.34}{100.33}{5}{61.38}{100.50}{6}
\emline{61.38}{100.50}{7}{63.17}{100.68}{8}
\emline{63.17}{100.68}{9}{64.71}{100.86}{10}
\emline{64.71}{100.86}{11}{65.99}{101.04}{12}
\emline{65.99}{101.04}{13}{67.03}{101.22}{14}
\emline{67.03}{101.22}{15}{67.81}{101.41}{16}
\emline{67.81}{101.41}{17}{68.34}{101.60}{18}
\emline{68.34}{101.60}{19}{68.62}{101.80}{20}
\emline{68.62}{101.80}{21}{68.65}{102.00}{22}
\emline{68.65}{102.00}{23}{68.42}{102.20}{24}
\emline{68.42}{102.20}{25}{67.95}{102.40}{26}
\emline{67.95}{102.40}{27}{67.22}{102.61}{28}
\emline{67.22}{102.61}{29}{66.24}{102.82}{30}
\emline{66.24}{102.82}{31}{65.01}{103.03}{32}
\emline{65.01}{103.03}{33}{63.53}{103.25}{34}
\emline{63.53}{103.25}{35}{61.80}{103.47}{36}
\emline{61.80}{103.47}{37}{59.81}{103.69}{38}
\emline{59.81}{103.69}{39}{55.09}{104.15}{40}
\emline{55.09}{104.15}{41}{48.67}{104.67}{42}
\emline{38.67}{104.67}{43}{32.83}{104.20}{44}
\emline{32.83}{104.20}{45}{27.99}{103.74}{46}
\emline{27.99}{103.74}{47}{25.95}{103.52}{48}
\emline{25.95}{103.52}{49}{24.16}{103.30}{50}
\emline{24.16}{103.30}{51}{22.62}{103.08}{52}
\emline{22.62}{103.08}{53}{21.34}{102.87}{54}
\emline{21.34}{102.87}{55}{20.31}{102.66}{56}
\emline{20.31}{102.66}{57}{19.52}{102.45}{58}
\emline{19.52}{102.45}{59}{18.99}{102.24}{60}
\emline{18.99}{102.24}{61}{18.71}{102.04}{62}
\emline{18.71}{102.04}{63}{18.69}{101.84}{64}
\emline{18.69}{101.84}{65}{18.91}{101.65}{66}
\emline{18.91}{101.65}{67}{19.39}{101.46}{68}
\emline{19.39}{101.46}{69}{20.11}{101.27}{70}
\emline{20.11}{101.27}{71}{21.09}{101.08}{72}
\emline{21.09}{101.08}{73}{22.32}{100.90}{74}
\emline{22.32}{100.90}{75}{23.80}{100.72}{76}
\emline{23.80}{100.72}{77}{25.54}{100.54}{78}
\emline{25.54}{100.54}{79}{27.52}{100.37}{80}
\emline{27.52}{100.37}{81}{32.25}{100.03}{82}
\emline{32.25}{100.03}{83}{38.67}{99.67}{84}
\emline{38.67}{104.67}{85}{41.17}{104.92}{86}
\emline{41.17}{104.92}{87}{46.17}{104.92}{88}
\emline{46.17}{104.92}{89}{48.67}{104.67}{90}
\emline{-6.33}{94.67}{91}{23.67}{109.67}{92}
\emline{23.67}{109.67}{93}{88.67}{109.67}{94}
\emline{88.67}{109.67}{95}{93.67}{109.67}{96}
\emline{93.67}{109.67}{97}{64.00}{94.67}{98}
\emline{64.00}{94.67}{99}{-6.33}{94.67}{100}
\emline{48.67}{59.67}{101}{54.51}{59.99}{102}
\emline{54.51}{59.99}{103}{59.34}{60.33}{104}
\emline{59.34}{60.33}{105}{61.38}{60.50}{106}
\emline{61.38}{60.50}{107}{63.17}{60.68}{108}
\emline{63.17}{60.68}{109}{64.71}{60.86}{110}
\emline{64.71}{60.86}{111}{65.99}{61.04}{112}
\emline{65.99}{61.04}{113}{67.03}{61.22}{114}
\emline{67.03}{61.22}{115}{67.81}{61.41}{116}
\emline{67.81}{61.41}{117}{68.34}{61.60}{118}
\emline{68.34}{61.60}{119}{68.62}{61.80}{120}
\emline{68.62}{61.80}{121}{68.65}{62.00}{122}
\emline{68.65}{62.00}{123}{68.42}{62.20}{124}
\emline{68.42}{62.20}{125}{67.95}{62.40}{126}
\emline{67.95}{62.40}{127}{67.22}{62.61}{128}
\emline{67.22}{62.61}{129}{66.24}{62.82}{130}
\emline{66.24}{62.82}{131}{65.01}{63.03}{132}
\emline{65.01}{63.03}{133}{63.53}{63.25}{134}
\emline{63.53}{63.25}{135}{61.80}{63.47}{136}
\emline{61.80}{63.47}{137}{59.81}{63.69}{138}
\emline{59.81}{63.69}{139}{55.09}{64.15}{140}
\emline{55.09}{64.15}{141}{48.67}{64.67}{142}
\emline{38.67}{64.67}{143}{32.83}{64.20}{144}
\emline{32.83}{64.20}{145}{27.99}{63.74}{146}
\emline{27.99}{63.74}{147}{25.95}{63.52}{148}
\emline{25.95}{63.52}{149}{24.16}{63.30}{150}
\emline{24.16}{63.30}{151}{22.62}{63.08}{152}
\emline{22.62}{63.08}{153}{21.34}{62.87}{154}
\emline{21.34}{62.87}{155}{20.31}{62.66}{156}
\emline{20.31}{62.66}{157}{19.52}{62.45}{158}
\emline{19.52}{62.45}{159}{18.99}{62.24}{160}
\emline{18.99}{62.24}{161}{18.71}{62.04}{162}
\emline{18.71}{62.04}{163}{18.69}{61.84}{164}
\emline{18.69}{61.84}{165}{18.91}{61.65}{166}
\emline{18.91}{61.65}{167}{19.39}{61.46}{168}
\emline{19.39}{61.46}{169}{20.11}{61.27}{170}
\emline{20.11}{61.27}{171}{21.09}{61.08}{172}
\emline{21.09}{61.08}{173}{22.32}{60.90}{174}
\emline{22.32}{60.90}{175}{23.80}{60.72}{176}
\emline{23.80}{60.72}{177}{25.54}{60.54}{178}
\emline{25.54}{60.54}{179}{27.52}{60.37}{180}
\emline{27.52}{60.37}{181}{32.25}{60.03}{182}
\emline{32.25}{60.03}{183}{38.67}{59.67}{184}
\emline{38.67}{64.67}{185}{41.17}{64.92}{186}
\emline{41.17}{64.92}{187}{46.17}{64.92}{188}
\emline{46.17}{64.92}{189}{48.67}{64.67}{190}
\emline{-6.33}{54.67}{191}{23.67}{69.67}{192}
\emline{88.67}{69.67}{193}{93.67}{69.67}{194}
\emline{93.67}{69.67}{195}{64.00}{54.67}{196}
\emline{64.00}{54.67}{197}{-6.33}{54.67}{198}
\emline{18.67}{102.00}{199}{20.46}{100.18}{200}
\emline{20.46}{100.18}{201}{22.08}{98.35}{202}
\emline{22.08}{98.35}{203}{23.51}{96.52}{204}
\emline{23.51}{96.52}{205}{24.78}{94.68}{206}
\emline{24.78}{94.68}{207}{25.86}{92.83}{208}
\emline{25.86}{92.83}{209}{26.77}{90.98}{210}
\emline{26.77}{90.98}{211}{27.50}{89.13}{212}
\emline{27.50}{89.13}{213}{28.05}{87.26}{214}
\emline{28.05}{87.26}{215}{28.43}{85.39}{216}
\emline{28.43}{85.39}{217}{28.63}{83.52}{218}
\emline{28.63}{83.52}{219}{28.65}{81.64}{220}
\emline{28.65}{81.64}{221}{28.50}{79.75}{222}
\emline{28.50}{79.75}{223}{28.17}{77.86}{224}
\emline{28.17}{77.86}{225}{27.66}{75.96}{226}
\emline{27.66}{75.96}{227}{26.98}{74.06}{228}
\emline{26.98}{74.06}{229}{26.12}{72.15}{230}
\emline{26.12}{72.15}{231}{25.08}{70.23}{232}
\emline{25.08}{70.23}{233}{23.86}{68.31}{234}
\emline{23.86}{68.31}{235}{22.47}{66.38}{236}
\emline{22.47}{66.38}{237}{20.90}{64.45}{238}
\emline{20.90}{64.45}{239}{18.67}{62.00}{240}
\emline{68.67}{61.67}{241}{66.87}{63.49}{242}
\emline{66.87}{63.49}{243}{65.26}{65.32}{244}
\emline{65.26}{65.32}{245}{63.83}{67.15}{246}
\emline{63.83}{67.15}{247}{62.57}{68.99}{248}
\emline{62.57}{68.99}{249}{61.49}{70.84}{250}
\emline{61.49}{70.84}{251}{60.59}{72.69}{252}
\emline{60.59}{72.69}{253}{59.87}{74.54}{254}
\emline{59.87}{74.54}{255}{59.33}{76.41}{256}
\emline{59.33}{76.41}{257}{58.96}{78.28}{258}
\emline{58.96}{78.28}{259}{58.77}{80.15}{260}
\emline{58.77}{80.15}{261}{58.77}{82.03}{262}
\emline{58.77}{82.03}{263}{58.94}{83.92}{264}
\emline{58.94}{83.92}{265}{59.29}{85.81}{266}
\emline{59.29}{85.81}{267}{59.81}{87.71}{268}
\emline{59.81}{87.71}{269}{60.52}{89.61}{270}
\emline{60.52}{89.61}{271}{61.40}{91.52}{272}
\emline{61.40}{91.52}{273}{62.47}{93.44}{274}
\emline{62.47}{93.44}{275}{63.71}{95.36}{276}
\emline{63.71}{95.36}{277}{65.13}{97.29}{278}
\emline{65.13}{97.29}{279}{66.73}{99.22}{280}
\emline{66.73}{99.22}{281}{69.00}{101.67}{282}
\emline{38.67}{99.67}{283}{41.17}{99.42}{284}
\emline{41.17}{99.42}{285}{46.17}{99.42}{286}
\emline{46.17}{99.42}{287}{48.67}{99.67}{288}
\emline{88.67}{69.67}{289}{62.34}{69.67}{290}
\emline{58.67}{69.67}{291}{53.67}{69.67}{292}
\emline{48.67}{69.67}{293}{44.00}{69.67}{294}
\emline{38.67}{69.67}{295}{34.00}{69.67}{296}
\emline{28.67}{69.67}{297}{24.67}{69.67}{298}
\emline{38.67}{59.67}{299}{41.17}{59.42}{300}
\emline{41.17}{59.42}{301}{46.17}{59.42}{302}
\emline{46.17}{59.42}{303}{48.67}{59.67}{304}
\put(74.67,105.67){\makebox(0,0)[cc]{\footnotesize $(n,1)$}}
\put(74.67,65.67){\makebox(0,0)[cc]{\footnotesize $(n,1)$}}
\put(44.67,83.67){\makebox(0,0)[cc]{\footnotesize $(1,0)$}}
\put(44.67,102.00){\makebox(0,0)[cc]{\tiny $(n-1,1)$}}
\put(44.67,62.00){\makebox(0,0)[cc]{\tiny $(n-1,1)$}}
\put(-8.33,0.00){}
\emline{30.00}{102.33}{305}{30.00}{102.33}{306}
\emline{131.00}{61.67}{307}{136.00}{61.67}{308}
\emline{140.33}{61.67}{309}{144.33}{61.67}{310}
\emline{138.00}{61.67}{311}{138.00}{61.67}{312}
\emline{146.33}{61.67}{313}{146.33}{61.67}{314}
\emline{148.33}{61.67}{315}{152.00}{61.67}{316}
\put(131.00,61.67){\circle*{0.67}}
\emline{131.00}{61.67}{317}{129.32}{63.15}{318}
\emline{129.32}{63.15}{319}{127.82}{64.84}{320}
\emline{127.82}{64.84}{321}{126.48}{66.76}{322}
\emline{126.48}{66.76}{323}{125.33}{68.89}{324}
\emline{125.33}{68.89}{325}{124.34}{71.23}{326}
\emline{124.34}{71.23}{327}{123.00}{76.00}{328}
\emline{130.67}{61.67}{329}{130.67}{66.00}{330}
\emline{130.67}{67.67}{331}{130.67}{67.67}{332}
\emline{130.67}{69.33}{333}{130.67}{73.00}{334}
\emline{130.67}{76.67}{335}{130.67}{80.67}{336}
\emline{130.67}{74.67}{337}{130.67}{74.67}{338}
\emline{130.67}{61.67}{339}{115.67}{61.67}{340}
\emline{130.67}{61.67}{341}{133.38}{61.45}{342}
\emline{133.38}{61.45}{343}{136.04}{61.20}{344}
\emline{136.04}{61.20}{345}{138.66}{60.92}{346}
\emline{138.66}{60.92}{347}{141.23}{60.60}{348}
\emline{141.23}{60.60}{349}{143.75}{60.25}{350}
\emline{143.75}{60.25}{351}{146.23}{59.87}{352}
\emline{146.23}{59.87}{353}{148.66}{59.45}{354}
\emline{148.66}{59.45}{355}{151.04}{59.00}{356}
\emline{151.04}{59.00}{357}{153.38}{58.52}{358}
\emline{153.38}{58.52}{359}{155.67}{58.00}{360}
\emline{125.67}{61.67}{361}{126.08}{63.92}{362}
\emline{126.08}{63.92}{363}{127.33}{65.33}{364}
\put(123.33,63.67){\makebox(0,0)[cc]{$\gamma$}}
\emline{150.00}{61.67}{365}{150.00}{59.33}{366}
\put(153.67,60.4){\makebox(0,0)[cc]{\small $\beta$}}
\put(120.67,79){\makebox(0,0)[cc]{\footnotesize $ (1,0)$}}
\put(118.33,58.67){\makebox(0,0)[cc]{\tiny $(n-1,1)$}}
\put(140.00,57.4){\makebox(0,0)[cc]{\footnotesize $(n,1)$}}
\put(41.33,43.00){\makebox(0,0)[cc]{(a)}}
\put(129.67,43.00){\makebox(0,0)[cc]{(b)}}
\end{picture}

\vspace{-4cm}

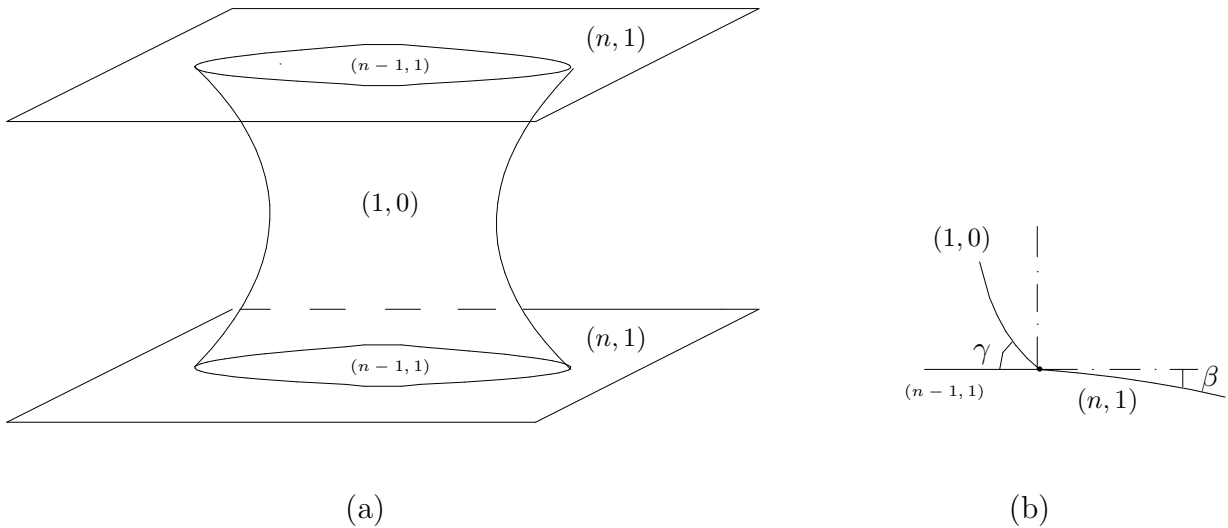
Fig.2: (a) Euclidean  brane configuration describing W-boson pair
production in 1+1 dimensional U(2) gauge theory. (b) $(p,q)$-strings' junction.

\vspace{0.8cm}

Hence we suggest the
following Euclidean brane configuration responsible for the
pair production in 1+1 dimensions (see Fig.2).
First of all, it should have rotational symmetry.
There are two parallel $(n,1)$-string worldvolumes which naively
look like   two planes.
Later on we will see that in fact their
shapes deviate logarithmically from the plane form at infinity.
This is a peculiar property of the two dimensional physics.
However, this can be interpreted as a  renormalization of the
W-boson mass and will not affect the form of the final answer
for the probability. For this reason we will first assume that
the $(n,1)$-string worldsheet is flat.
Two  additional ingredients of the picture are:
minimal surface formed by the  F-string
stretched between two $(n,1)$-string worldsheets and two
discs of the radius $R$ representing the worldsheets
of $(n-1,1)$-strings. The reason why these worldsheets
must be flat is obvious from the physical point of view since
the electric force is acting only on its boundaries.

We choose cylindrical coordinates $(r,\phi, z)$ such that $(n-1,1)$-string
surface is orthogonal to the $z$ axis.
The angle $\gamma$ between the minimal F-surface and the
$(n-1,1)$-string worldsheet can be determined
geometrically from the junction condition.
We suppose that $n$ or $1/g_{st}$ is large enough,
then~\footnote{The condition $\cos\gamma=E$ seems to be universal for
the problem at hand -- as we will see later,
it holds in AdS space too.}

\be
\label{gamma}
\cos\gamma = \frac{ng_{st}}{\sqrt{1+ (ng_{st})^2}}
\ee
The angle $\b$ between $(n,1)$ and $(n-1,1)$-string worldsheets
in our case is always small since it is given by

\be
\sin \b=\frac{1}{g_{st}(n^2+1/g_{st}^2)}
\ee
It follows from (\ref{gamma}) that when
the coupling constant is small, F string surface is orthogonal
to $(n-1,1)$ string worldsheet and the minimal surface becomes a cylinder.
At the large coupling F-string should join the dyonic
string at zero angle.

Let us briefly comment on how the process looks from the  Euclidean time evolution
point of view. First,
at some point in the Euclidean time two
finite pieces of the strings leave D-branes.
Later on they touch in the bulk and  rearrange
into the two strings stretched between the two D-branes.
After the materialization in Minkowski space
they move apart.
This evolution process is presented in Fig.3.

\vspace{1cm}

\special{em:linewidth 0.4pt}
\unitlength 1.00mm
\linethickness{0.4pt}
\begin{picture}(140.67,70.33)
\emline{10.33}{49.67}{1}{40.33}{49.67}{2}
\emline{60.33}{49.67}{3}{90.33}{49.67}{4}
\emline{110.00}{49.67}{5}{140.33}{49.67}{6}
\emline{10.67}{70.33}{7}{40.67}{70.33}{8}
\emline{60.67}{70.33}{9}{90.67}{70.33}{10}
\emline{110.33}{70.33}{11}{140.67}{70.33}{12}
\emline{15.00}{70.33}{13}{16.28}{68.33}{14}
\emline{16.28}{68.33}{15}{17.56}{66.61}{16}
\emline{17.56}{66.61}{17}{18.85}{65.15}{18}
\emline{18.85}{65.15}{19}{20.13}{63.98}{20}
\emline{20.13}{63.98}{21}{21.41}{63.07}{22}
\emline{21.41}{63.07}{23}{22.69}{62.44}{24}
\emline{22.69}{62.44}{25}{23.97}{62.09}{26}
\emline{23.97}{62.09}{27}{25.26}{62.01}{28}
\emline{25.26}{62.01}{29}{26.54}{62.20}{30}
\emline{26.54}{62.20}{31}{27.82}{62.66}{32}
\emline{27.82}{62.66}{33}{29.10}{63.40}{34}
\emline{29.10}{63.40}{35}{30.38}{64.42}{36}
\emline{30.38}{64.42}{37}{31.67}{65.70}{38}
\emline{31.67}{65.70}{39}{32.95}{67.26}{40}
\emline{32.95}{67.26}{41}{35.00}{70.33}{42}
\emline{15.33}{49.67}{43}{16.57}{51.67}{44}
\emline{16.57}{51.67}{45}{17.81}{53.39}{46}
\emline{17.81}{53.39}{47}{19.05}{54.85}{48}
\emline{19.05}{54.85}{49}{20.29}{56.02}{50}
\emline{20.29}{56.02}{51}{21.53}{56.93}{52}
\emline{21.53}{56.93}{53}{22.77}{57.56}{54}
\emline{22.77}{57.56}{55}{24.01}{57.91}{56}
\emline{24.01}{57.91}{57}{25.25}{57.99}{58}
\emline{25.25}{57.99}{59}{26.49}{57.80}{60}
\emline{26.49}{57.80}{61}{27.73}{57.34}{62}
\emline{27.73}{57.34}{63}{28.97}{56.60}{64}
\emline{28.97}{56.60}{65}{30.21}{55.58}{66}
\emline{30.21}{55.58}{67}{31.45}{54.30}{68}
\emline{31.45}{54.30}{69}{32.69}{52.74}{70}
\emline{32.69}{52.74}{71}{34.67}{49.67}{72}
\put(24.67,40.00){\makebox(0,0)[cc]{(a)}}
\put(74.67,40.00){\makebox(0,0)[cc]{(b)}}
\put(124.67,40.00){\makebox(0,0)[cc]{(c)}}
\emline{115.00}{70.33}{73}{116.97}{68.97}{74}
\emline{116.97}{68.97}{75}{118.66}{67.61}{76}
\emline{118.66}{67.61}{77}{120.07}{66.25}{78}
\emline{120.07}{66.25}{79}{121.20}{64.89}{80}
\emline{121.20}{64.89}{81}{122.06}{63.53}{82}
\emline{122.06}{63.53}{83}{122.65}{62.17}{84}
\emline{122.65}{62.17}{85}{122.95}{60.82}{86}
\emline{122.95}{60.82}{87}{122.98}{59.46}{88}
\emline{122.98}{59.46}{89}{122.73}{58.10}{90}
\emline{122.73}{58.10}{91}{122.20}{56.74}{92}
\emline{122.20}{56.74}{93}{121.40}{55.38}{94}
\emline{121.40}{55.38}{95}{120.32}{54.02}{96}
\emline{120.32}{54.02}{97}{118.96}{52.66}{98}
\emline{118.96}{52.66}{99}{117.33}{51.30}{100}
\emline{117.33}{51.30}{101}{115.00}{49.67}{102}
\emline{135.00}{70.33}{103}{133.03}{68.97}{104}
\emline{133.03}{68.97}{105}{131.34}{67.61}{106}
\emline{131.34}{67.61}{107}{129.93}{66.25}{108}
\emline{129.93}{66.25}{109}{128.80}{64.89}{110}
\emline{128.80}{64.89}{111}{127.94}{63.53}{112}
\emline{127.94}{63.53}{113}{127.35}{62.17}{114}
\emline{127.35}{62.17}{115}{127.05}{60.82}{116}
\emline{127.05}{60.82}{117}{127.02}{59.46}{118}
\emline{127.02}{59.46}{119}{127.27}{58.10}{120}
\emline{127.27}{58.10}{121}{127.80}{56.74}{122}
\emline{127.80}{56.74}{123}{128.60}{55.38}{124}
\emline{128.60}{55.38}{125}{129.68}{54.02}{126}
\emline{129.68}{54.02}{127}{131.04}{52.66}{128}
\emline{131.04}{52.66}{129}{132.67}{51.30}{130}
\emline{132.67}{51.30}{131}{135.00}{49.67}{132}
\emline{65.00}{70.33}{133}{66.06}{68.18}{134}
\emline{66.06}{68.18}{135}{67.13}{66.27}{136}
\emline{67.13}{66.27}{137}{68.19}{64.61}{138}
\emline{68.19}{64.61}{139}{69.26}{63.18}{140}
\emline{69.26}{63.18}{141}{70.32}{62.00}{142}
\emline{70.32}{62.00}{143}{71.38}{61.06}{144}
\emline{71.38}{61.06}{145}{72.45}{60.36}{146}
\emline{72.45}{60.36}{147}{73.51}{59.90}{148}
\emline{73.51}{59.90}{149}{74.57}{59.68}{150}
\emline{74.57}{59.68}{151}{75.64}{59.71}{152}
\emline{75.64}{59.71}{153}{76.70}{59.97}{154}
\emline{76.70}{59.97}{155}{77.77}{60.48}{156}
\emline{77.77}{60.48}{157}{78.83}{61.23}{158}
\emline{78.83}{61.23}{159}{79.89}{62.22}{160}
\emline{79.89}{62.22}{161}{80.96}{63.45}{162}
\emline{80.96}{63.45}{163}{82.02}{64.92}{164}
\emline{82.02}{64.92}{165}{83.09}{66.64}{166}
\emline{83.09}{66.64}{167}{85.00}{70.33}{168}
\emline{65.00}{49.67}{169}{66.11}{51.77}{170}
\emline{66.11}{51.77}{171}{67.22}{53.62}{172}
\emline{67.22}{53.62}{173}{68.33}{55.22}{174}
\emline{68.33}{55.22}{175}{69.44}{56.58}{176}
\emline{69.44}{56.58}{177}{70.56}{57.69}{178}
\emline{70.56}{57.69}{179}{71.67}{58.56}{180}
\emline{71.67}{58.56}{181}{72.78}{59.17}{182}
\emline{72.78}{59.17}{183}{73.89}{59.54}{184}
\emline{73.89}{59.54}{185}{76.11}{59.54}{186}
\emline{76.11}{59.54}{187}{77.22}{59.17}{188}
\emline{77.22}{59.17}{189}{78.33}{58.56}{190}
\emline{78.33}{58.56}{191}{79.44}{57.69}{192}
\emline{79.44}{57.69}{193}{80.56}{56.58}{194}
\emline{80.56}{56.58}{195}{81.67}{55.22}{196}
\emline{81.67}{55.22}{197}{82.78}{53.62}{198}
\emline{82.78}{53.62}{199}{83.89}{51.77}{200}
\emline{83.89}{51.77}{201}{85.00}{49.67}{202}
\end{picture}

\vspace{-3.5cm}

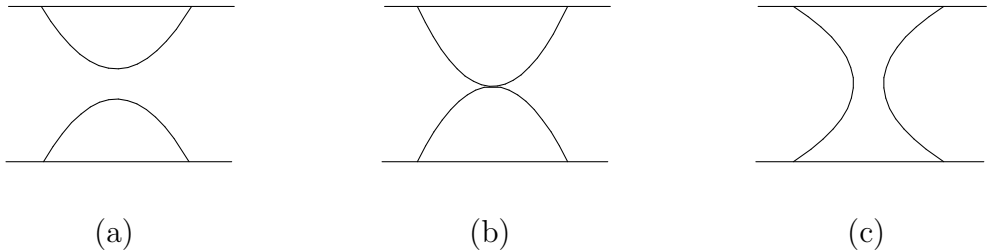
Fig.3: Euclidean time slices of the strings evolution process in the constant
electric field.

\vspace{1cm}

\noindent
To get the rate of the pair production
we have to find the minimum of the following action:

\be
\label{1+1action}
S=T_{1,0}A_{1,0}+2(T_{n-1,1}-T_{n,1})\pi R^2
\ee
where $R$ is a boundary radius of the F string
worldsheet, $T_{p,q}=\sqrt{p^2+q^2/g_{str}^2}$ is the
$(p,q)$-string's tension  and  $A_{p,q}$ is the  area of the
$(p,q)$-string's worldsheet:

\be
A_{p,q}=2\pi \int  r(z)\sqrt{1+(r'(z))^2}dz
\ee
The action $A_{p,q}$ has the following integral of motion:

\be
\label{integral}
\frac{r(z)}{\sqrt{1+(r'(z))^2}}=\const
\ee
For the F-string there is a minimal radius $r=r_0$ where
$r'_0=0$ and therefore $\const=r_0$.
The equation of the minimal surface  is

\be
r(z)=r_0 \cosh \frac{z-v/2}{r_0}
\ee
Here $z=0$ and $z=v$ are coordinates of the two dyonic branes.
Therefore,

\be
A_{1,0}=\pi r_0 v+2\pi R\sqrt{R^2-r_0^2}
\ee
Since the condition (\ref{gamma}) implies

\be
\label{derivative}
\left. \frac{dr}{dz} \right|_{z=v}=\frac{E}{\sqrt{1-E^2}}
\ee
we can express the minimal and the maximal radii of the surface as

\be
r_0=\frac{v}{\log\frac{1+E}{1-E}}, \qquad
R=\frac{v}{\sqrt{1-E^2} \log\frac{1+E}{1-E}}
\ee
It is clear that the critical electric field
in our approach has quite geometrical interpretation.
It is the field when $r_0$ vanishes and the minimal
surface becomes disconnected. Note that
we have no Gross-Ooguri phase transition
in our problem at any subcritical electric field.
When the electrical field tends to the critical value, $R \rightarrow
\infty$. This means that effectively the infinitely long string is produced.

Thus, the minimum of the  action~(\ref{1+1action}) is given by:

\be
S_{min}=\frac{\pi v^2}{\log\frac{1+E}{1-E}}
\ee
Finally, after restoring the $\a'$ and $g$ dependence
we obtain the exponential factor

\be
\label{rate2dim}
w \propto \exp\left(-\frac{2 g^2 \pi^2 \a' v^2}
{\log\frac{1+2 \pi \a' g E}{1-2 \pi \a' g E}} \right)
\ee
which agrees with the Bachas-Porrati result in the
following sense: our boundary conditions select
the single (ground) state from the whole tower of the stringy modes.
That is why our result coincides with the
leading term from the Bachas-Porrati probability series
where all string states are involved.
Note that from the field theory viewpoint the effective
charge of the W-boson is $2g$ rather then $g$ since it is
in the adjoint representation of the gauge group.

The formulae (\ref{rate2dim}) yields the  exponential factor in the
leading term in the full stringy probability (which would be a sum of the type
(2.6)). Let us try to understand how a double summation would arise
in our framework. The sum over $S$ (string states) corresponds to the more
complicated extremal surfaces without rotational symmetry.
The sum over $n$ corresponds to the "multi-instanton" contributions, which in our
case look like clusters  of $n$  extremal surfaces touching each other on the boundary branes.
Preexponential factors can be restored from the
consideration of the determinant of the fluctuations around the minimal surface.

{\it Deformation of the worldsheet in 1+1 dimensional case}

Now let us take into account the effects coming from
the deformation of the $(n,1)$-string worldsheets.
We introduce the radius $R_{\infty} \gg R$ for the boundaries
of this worldsheets and the distance $v_{\infty}$
between them at $R_{\infty}$. In a moment
it will be clear that $R_{\infty}$ serves as a IR regulator scale.
The action to be minimized is given by
\be
S=T_{1,0}A_{1,0}+2(T_{n-1,1}-T_{n,1})\pi R^2+2T_{n,1}(A_{n,1}-\pi R_{\infty}^2)
\ee
The  $(n-1,1)$-string surface has the form:

\be
r(z)=R\sin\b \cosh \frac{z+z_0}{R\sin\b}
\ee
where the constant $z_0$ is determined by the relation

\be
R_{\infty}=R\sin\b \cosh \frac{z+v_{\infty}/2}{R\sin\b}
\ee
We have:

\be
A_{n,1}-\pi R_{\infty}^2=\pi Rv_{\infty} \sin\b  -
2\pi R^2+O(e^{-R/R_{\infty}})
\ee
Therefore,

\be
S_{min}=\frac{\pi v^2(1+v_{\infty}/v)}{\log\frac{1+E}{1-E}}
\ee
Since we can rewrite  $v_{\infty}/v$ as follows:

\be
{v_{\infty} \over v} =  2 g \frac{\sqrt{1-E^2} }{\log\frac{1+E}{1-E}}
\log \frac{2 R_{\infty}}{R \sin \b},
\ee
we see that  deformation of the
dyonic string surface results in additional term in the effective action,
which is proportional to the coupling constant $g$.
Such "quantum" terms are beyond our approximation here, since
their appearance is natural in  preexponential ("quantum determinant")
factors rather than in exponential factors. Note that from the
1+1 dimensional field theory viewpoint
this term  can be interpreted as a renormalization of the
scalar vacuum value~$v$.

{\it Pair production in higher dimensions}

To fix the nontrivial field
on the brane in higher dimensions, say on the D3-brane in d=4, one has to
consider the bound state of the D3-brane with $N$ F1-strings. The number $N$
of the F1-branes per unit transverse volume fixes the
electric field on the D3-brane worldvolume as follows
\be
\frac{N}{V_t}= \frac {E}{\alpha^{'}g_{str}\sqrt {E_{cr}^{2}-E^2}}
\ee
The pair creation corresponds to the process when one F1-string, initially
bound to the one D3-brane, finally becomes stretched between two D3-branes.
The calculation of the probability goes similarly to the 1+1 dimensional case
but instead of the junction smooth F1-string worldvolume has to be
considered. The effective action providing the production amplitude can be
derived in the similar manner and involves  the surface term and the electric
field contribution. This gives the same exponential factors (\ref{rate2dim})
as in the 1+1 dimensional case. The differences in higher dimensions are twofold. First,
asymptotically there is no back reaction from the escaping F1-string and
therefore there is no logarithmic  renormalization of the scalar condensate.
Second, there are different preexponential factors in different dimensions
which however are beyond our approximation.

\subsection{Production of the fundamental matter}

To consider the production of the pair of
particles in the fundamental representation we represent the
fundamental matter by the
set of branes. In this section for the simplicity
U(1) gauge group  will be considered. One should
distinguish several realizations of the fundamental
matter in the different versions of the string theory.

Let us start with IIB/F theory picture. In the simplest
IIB realization of the  N=4 SUSY theory
matter is represented
by the fundamental strings connecting the D3 brane
where the gauge theory is defined on and
$N_{F}$ D5 branes. In the F theory
it is represented by the strings connecting D7 branes
wrapped around the torus in $(10,11)$ dimensions
and D3 branes. The
orientation of the branes, say in F-theory, is as follows:  worldvolume
coordinates of the D3 brane are $(x_0,x_1,x_2,x_3)$ and
of the D7 are $(x_0,x_1,x_2,x_3,x_7,x_8,x_{10},x_{11})$.
The coordinates of the prime interest are $(x_5,x_6)$ since D3 brane is
localized at the origin while $k$-th D7 is localized
at $z_k=x_{5,k}+ix_{6,k}=m_k$, where $m_k$
is the mass of the k-th flavor. It is assumed that all D3 and D7
branes are localized at the origin in $(x_4,x_9)$ directions.

\vspace{0.8cm}

\special{em:linewidth 0.4pt}
\unitlength 1.00mm
\linethickness{0.4pt}
\begin{picture}(123.67,86.00)
\emline{48.67}{59.67}{1}{54.51}{59.99}{2}
\emline{54.51}{59.99}{3}{59.35}{60.33}{4}
\emline{59.35}{60.33}{5}{61.39}{60.50}{6}
\emline{61.39}{60.50}{7}{63.17}{60.68}{8}
\emline{63.17}{60.68}{9}{64.71}{60.86}{10}
\emline{64.71}{60.86}{11}{66.00}{61.04}{12}
\emline{66.00}{61.04}{13}{67.03}{61.22}{14}
\emline{67.03}{61.22}{15}{67.81}{61.41}{16}
\emline{67.81}{61.41}{17}{68.34}{61.60}{18}
\emline{68.34}{61.60}{19}{68.62}{61.80}{20}
\emline{68.62}{61.80}{21}{68.65}{62.00}{22}
\emline{68.65}{62.00}{23}{68.43}{62.20}{24}
\emline{68.43}{62.20}{25}{67.95}{62.40}{26}
\emline{67.95}{62.40}{27}{67.22}{62.61}{28}
\emline{67.22}{62.61}{29}{66.25}{62.82}{30}
\emline{66.25}{62.82}{31}{65.02}{63.03}{32}
\emline{65.02}{63.03}{33}{63.53}{63.25}{34}
\emline{63.53}{63.25}{35}{61.80}{63.47}{36}
\emline{61.80}{63.47}{37}{59.82}{63.69}{38}
\emline{59.82}{63.69}{39}{55.09}{64.15}{40}
\emline{55.09}{64.15}{41}{48.67}{64.67}{42}
\emline{38.67}{64.67}{43}{32.83}{64.20}{44}
\emline{32.83}{64.20}{45}{27.99}{63.74}{46}
\emline{27.99}{63.74}{47}{25.95}{63.52}{48}
\emline{25.95}{63.52}{49}{24.17}{63.30}{50}
\emline{24.17}{63.30}{51}{22.63}{63.08}{52}
\emline{22.63}{63.08}{53}{21.34}{62.87}{54}
\emline{21.34}{62.87}{55}{20.31}{62.66}{56}
\emline{20.31}{62.66}{57}{19.53}{62.45}{58}
\emline{19.53}{62.45}{59}{19.00}{62.24}{60}
\emline{19.00}{62.24}{61}{18.72}{62.04}{62}
\emline{18.72}{62.04}{63}{18.69}{61.84}{64}
\emline{18.69}{61.84}{65}{18.91}{61.65}{66}
\emline{18.91}{61.65}{67}{19.39}{61.46}{68}
\emline{19.39}{61.46}{69}{20.12}{61.27}{70}
\emline{20.12}{61.27}{71}{21.09}{61.08}{72}
\emline{21.09}{61.08}{73}{22.32}{60.90}{74}
\emline{22.32}{60.90}{75}{23.81}{60.72}{76}
\emline{23.81}{60.72}{77}{25.54}{60.54}{78}
\emline{25.54}{60.54}{79}{27.52}{60.37}{80}
\emline{27.52}{60.37}{81}{32.25}{60.03}{82}
\emline{32.25}{60.03}{83}{38.67}{59.67}{84}
\emline{38.67}{64.67}{85}{41.17}{64.92}{86}
\emline{41.17}{64.92}{87}{46.17}{64.92}{88}
\emline{46.17}{64.92}{89}{48.67}{64.67}{90}
\emline{-6.33}{54.67}{91}{23.67}{69.67}{92}
\emline{88.67}{69.67}{93}{93.67}{69.67}{94}
\emline{93.67}{69.67}{95}{64.00}{54.67}{96}
\emline{64.00}{54.67}{97}{-6.33}{54.67}{98}
\emline{88.67}{69.67}{99}{62.34}{69.67}{100}
\emline{58.67}{69.67}{101}{53.67}{69.67}{102}
\emline{48.67}{69.67}{103}{44.00}{69.67}{104}
\emline{38.67}{69.67}{105}{34.00}{69.67}{106}
\emline{28.67}{69.67}{107}{24.67}{69.67}{108}
\emline{38.67}{59.67}{109}{41.17}{59.42}{110}
\emline{41.17}{59.42}{111}{46.17}{59.42}{112}
\emline{46.17}{59.42}{113}{48.67}{59.67}{114}
\put(-8.33,0.00){}
\emline{68.67}{61.67}{115}{66.81}{63.18}{116}
\emline{66.81}{63.18}{117}{65.14}{64.80}{118}
\emline{65.14}{64.80}{119}{63.65}{66.55}{120}
\emline{63.65}{66.55}{121}{62.36}{68.41}{122}
\emline{62.36}{68.41}{123}{61.25}{70.39}{124}
\emline{61.25}{70.39}{125}{60.33}{72.49}{126}
\emline{60.33}{72.49}{127}{59.61}{74.71}{128}
\emline{59.61}{74.71}{129}{59.07}{77.04}{130}
\emline{59.07}{77.04}{131}{58.67}{80.00}{132}
\emline{18.67}{61.67}{133}{20.35}{63.38}{134}
\emline{20.35}{63.38}{135}{21.89}{65.19}{136}
\emline{21.89}{65.19}{137}{23.31}{67.09}{138}
\emline{23.31}{67.09}{139}{24.60}{69.09}{140}
\emline{24.60}{69.09}{141}{25.76}{71.18}{142}
\emline{25.76}{71.18}{143}{26.80}{73.37}{144}
\emline{26.80}{73.37}{145}{27.70}{75.65}{146}
\emline{27.70}{75.65}{147}{29.00}{80.00}{148}
\emline{58.67}{80.00}{149}{58.33}{83.33}{150}
\emline{29.00}{80.00}{151}{29.67}{83.33}{152}
\emline{41.00}{81.67}{155}{38.67}{81.92}{156}
\emline{38.67}{81.92}{157}{36.61}{82.17}{158}
\emline{36.61}{82.17}{159}{34.82}{82.42}{160}
\emline{34.82}{82.42}{161}{33.30}{82.67}{162}
\emline{33.30}{82.67}{163}{32.05}{82.91}{164}
\emline{32.05}{82.91}{165}{31.07}{83.15}{166}
\emline{31.07}{83.15}{167}{30.36}{83.40}{168}
\emline{30.36}{83.40}{169}{29.92}{83.63}{170}
\emline{29.92}{83.63}{171}{29.75}{83.87}{172}
\emline{29.75}{83.87}{173}{29.85}{84.11}{174}
\emline{29.85}{84.11}{175}{30.22}{84.34}{176}
\emline{30.22}{84.34}{177}{30.86}{84.57}{178}
\emline{30.86}{84.57}{179}{31.76}{84.80}{180}
\emline{31.76}{84.80}{181}{32.94}{85.02}{182}
\emline{32.94}{85.02}{183}{34.39}{85.25}{184}
\emline{34.39}{85.25}{185}{36.11}{85.47}{186}
\emline{36.11}{85.47}{187}{38.09}{85.69}{188}
\emline{38.09}{85.69}{189}{41.33}{86.00}{190}
\emline{47.00}{81.33}{191}{49.36}{81.60}{192}
\emline{49.36}{81.60}{193}{51.43}{81.86}{194}
\emline{51.43}{81.86}{195}{53.22}{82.13}{196}
\emline{53.22}{82.13}{197}{54.73}{82.39}{198}
\emline{54.73}{82.39}{199}{55.95}{82.66}{200}
\emline{55.95}{82.66}{201}{56.89}{82.92}{202}
\emline{56.89}{82.92}{203}{57.54}{83.19}{204}
\emline{57.54}{83.19}{205}{57.91}{83.45}{206}
\emline{57.91}{83.45}{207}{57.99}{83.72}{208}
\emline{57.99}{83.72}{209}{57.80}{83.98}{210}
\emline{57.80}{83.98}{211}{57.31}{84.25}{212}
\emline{57.31}{84.25}{213}{56.55}{84.52}{214}
\emline{56.55}{84.52}{215}{55.49}{84.78}{216}
\emline{55.49}{84.78}{217}{54.16}{85.05}{218}
\emline{54.16}{85.05}{219}{52.54}{85.31}{220}
\emline{52.54}{85.31}{221}{50.64}{85.58}{222}
\emline{50.64}{85.58}{223}{47.00}{86.00}{224}
\put(75.00,65.67){\makebox(0,0)[cc]{D3}}
\put(42.67,73.33){\makebox(0,0)[cc]{F1}}
\emline{41.00}{81.67}{225}{47.00}{81.33}{226}
\emline{42.00}{86.00}{227}{47.00}{86.00}{228}
\end{picture}

\vspace{-5.2cm}

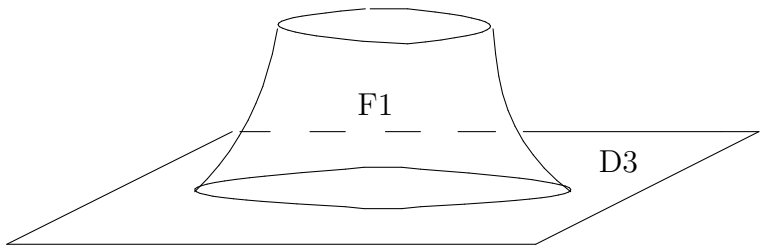
Fig.4: Brane configuration describing production of the fundamental matter.

\vspace{0.8cm}

Now let us identify the coordinates relevant for the
Euclidean configuration describing the Schwinger process.
Two coordinates are evident - these are the time and coordinate in the direction
of the external field. The metric on the   $(x_5,x_6)$
plane is nontrivial since D7 brane amounts to the cosmic
string metric around \cite{sen}.
However, the  cosmic string metric locally can be brought into the
flat form and the strings connecting D3 and D7 branes
become the straight lines. That is why the relevant
geometry of the Euclidean solution involves
the minimal  surface in $\R^3$.

However, there is some difference compared to the adjoint matter production case.
The point is that there is
no field on the worldvolume of D7 branes.
Therefore the minimal surface between the D3 and D7 branes
has to join D7 brane at the right angle. To get
such configuration we can take the minimal
surface relevant for the adjoint case and cut
it at $r=r_{min}$. It is evident that the
effective action in this case is twice less then in the adjoint case,
in agreement with the field theory result.
Hence the exponential factor in the production rate
of the fundamentals is given by

\be
w\propto \exp\left(-\frac{\pi m^2}{gE}\right)
\ee
The picture in IIA/M theory looks as follows. In IIA picture
one has D4 branes and $N_F$ D6-branes located at
positions $m_i$ in $z$ direction.
Fundamental matter is represented by the
strings connecting D6 and D4-branes. D6-branes amount to
the nontrivial Taub-NUT metric around, and
therefore one has to consider the minimal surface
taking into account this metric. In M-theory picture the
field theory is defined  on the worldvolume of the M5-brane
embedded into $d=11$ manifold whose four coordinates are
endowed with Taub-Nut metric. One can derive the
same result as in IIB case considering the minimal
surface in the Taub-Nut metric.

\vspace{-0.1cm}

\subsection{Monopole pair production}

Let us discuss the monopole pair production.
To this aim we have to consider stretched
D1-branes instead of the F1-branes.The
configuration looks similar to the charge case.
The probability of the monopole pair production
in the field theory limit is

\be
w \propto exp(-\pi m_{mon}^2/g_mB)
\ee
where $m_{mon}=v/g$ is the monopole mass and
$g_m$ is the magnetic coupling obeying the
quantization condition $g g_m =2\pi$.
Calculation of the stringy deformed
probability amounts to

\be
w \propto \exp\left(-\frac{ \a' v^2}
{\log\frac{1+g \a'  B}{1- g \a'  B}} \right)
\ee
Note that there is a kind of S-duality
transformation relating charge and magnetic
probabilities. Moreover if there are both electric
and magnetic fields, the dyons could be produced.

Let us emphasize that similarly to the electric field
case there exists the critical magnetic field value,
such that the process of the monopole pair creation in this
magnetic field becomes unsuppressed. In the brane picture it
corresponds to the field value such that the minimal radius $r_0$
of the sutface vanishes.
Note that in spite of
the smallness of the monopole
pair production rate compared to
the electric pair production rate
the dependencies of the critical electric
and magnetic fields on the coupling constant
are the same

\be
E_{cr} =(2 \pi g)^{-1} \a'^{-1}
\ee
\vspace{-0.3cm}
\be
\label{Bcrit}
B_{cr} = g^{-1} \a'^{-1}
\ee
In principle one can expect that the probability of the
monopole production
amounts from the annulus partition
function of the D string in the
magnetic background. The probability
calculated above is the contribution to the
total probability from the single string state.
However the lack of the perturbative description
of the D string makes the naive direct calculation
of the partition function impossible and we have
no  formulae from the string theory side to compare with.

The analysis above concerns the 3+1 dimensional
theory but there is some funny counterpart in
1+1 dimensions. If we consider the theory on the
worldvolume of $(p,q)$ string at large $p$ then the
process of escaping, say $(1,k)$ string due to the junction
is possible. Since $p$ is assumed to be large, the
backreaction of the junction on the $(p,q)$ string
can be neglected. However since there are no monopoles
in the two dimensions the interpretation of the
tunneling process is different. Namely, it corresponds
to the nonperturbative change of the rank of the
gauge group locally. The process looks as follows.
First, the small region where the rank of the gauge
decreases by one emerges
in the Euclidean space. Then it starts to expand in
the Minkowski space-time and finally the rank
of the gauge group decreases along the whole 1+1 dimensional space-time.

\vspace{-0.2cm}

\subsection{Pair production at finite temperature}

Let us consider the periodical Euclidean configuration
in the field theory limit. Consider first the situation
when $\beta = 1 / T$ is larger then the boundary cylinder
radius (Fig.5a). In this case naively there is no $T$ dependence
in the amplitude. When $\beta$ approaches the
radius of the cylinder the configuration changes into
the "fish" one (Fig.5b).
In the field theory limit we have
the periodic array of the
of deformed cylinders (Fig.5c).

\vspace{0.9cm}

\special{em:linewidth 0.4pt}
\unitlength 1.00mm
\linethickness{0.4pt}
\begin{picture}(142.67,108.33)
\emline{7.01}{99.00}{1}{5.03}{99.54}{2}
\emline{5.03}{99.54}{3}{4.09}{100.05}{4}
\emline{4.09}{100.05}{5}{4.19}{100.53}{6}
\emline{4.19}{100.53}{7}{7.01}{101.33}{8}
\emline{9.01}{101.33}{9}{10.99}{100.90}{10}
\emline{10.99}{100.90}{11}{11.92}{100.44}{12}
\emline{11.92}{100.44}{13}{11.82}{99.95}{14}
\emline{11.82}{99.95}{15}{9.01}{99.00}{16}
\emline{7.01}{99.00}{17}{9.01}{99.00}{18}
\emline{7.01}{101.33}{19}{9.01}{101.33}{20}
\emline{7.01}{86.00}{21}{5.03}{86.54}{22}
\emline{5.03}{86.54}{23}{4.09}{87.05}{24}
\emline{4.09}{87.05}{25}{4.19}{87.53}{26}
\emline{4.19}{87.53}{27}{7.01}{88.33}{28}
\emline{9.01}{88.33}{29}{10.99}{87.90}{30}
\emline{10.99}{87.90}{31}{11.92}{87.44}{32}
\emline{11.92}{87.44}{33}{11.82}{86.95}{34}
\emline{11.82}{86.95}{35}{9.01}{86.00}{36}
\emline{7.01}{86.00}{37}{9.01}{86.00}{38}
\emline{7.01}{88.33}{39}{9.01}{88.33}{40}
\emline{4.34}{99.67}{41}{5.21}{97.41}{42}
\emline{5.21}{97.41}{43}{5.67}{95.15}{44}
\emline{5.67}{95.15}{45}{5.73}{92.88}{46}
\emline{5.73}{92.88}{47}{5.39}{90.62}{48}
\emline{5.39}{90.62}{49}{4.00}{87.00}{50}
\emline{12.01}{100.00}{51}{11.03}{97.74}{52}
\emline{11.03}{97.74}{53}{10.47}{95.48}{54}
\emline{10.47}{95.48}{55}{10.35}{93.21}{56}
\emline{10.35}{93.21}{57}{10.64}{90.95}{58}
\emline{10.64}{90.95}{59}{12.01}{87.33}{60}
\emline{18.34}{99.00}{61}{16.36}{99.54}{62}
\emline{16.36}{99.54}{63}{15.42}{100.05}{64}
\emline{15.42}{100.05}{65}{15.52}{100.53}{66}
\emline{15.52}{100.53}{67}{18.34}{101.33}{68}
\emline{20.34}{101.33}{69}{22.32}{100.90}{70}
\emline{22.32}{100.90}{71}{23.25}{100.44}{72}
\emline{23.25}{100.44}{73}{23.15}{99.95}{74}
\emline{23.15}{99.95}{75}{20.34}{99.00}{76}
\emline{18.34}{99.00}{77}{20.34}{99.00}{78}
\emline{18.34}{101.33}{79}{20.34}{101.33}{80}
\emline{18.34}{86.00}{81}{16.36}{86.54}{82}
\emline{16.36}{86.54}{83}{15.42}{87.05}{84}
\emline{15.42}{87.05}{85}{15.52}{87.53}{86}
\emline{15.52}{87.53}{87}{18.34}{88.33}{88}
\emline{20.34}{88.33}{89}{22.32}{87.90}{90}
\emline{22.32}{87.90}{91}{23.25}{87.44}{92}
\emline{23.25}{87.44}{93}{23.15}{86.95}{94}
\emline{23.15}{86.95}{95}{20.34}{86.00}{96}
\emline{18.34}{86.00}{97}{20.34}{86.00}{98}
\emline{18.34}{88.33}{99}{20.34}{88.33}{100}
\emline{15.67}{100.00}{101}{16.54}{97.74}{102}
\emline{16.54}{97.74}{103}{17.00}{95.48}{104}
\emline{17.00}{95.48}{105}{17.06}{93.21}{106}
\emline{17.06}{93.21}{107}{16.72}{90.95}{108}
\emline{16.72}{90.95}{109}{15.34}{87.33}{110}
\emline{23.34}{100.00}{111}{22.36}{97.74}{112}
\emline{22.36}{97.74}{113}{21.81}{95.48}{114}
\emline{21.81}{95.48}{115}{21.68}{93.21}{116}
\emline{21.68}{93.21}{117}{21.98}{90.95}{118}
\emline{21.98}{90.95}{119}{23.34}{87.33}{120}
\emline{29.67}{99.00}{121}{27.69}{99.54}{122}
\emline{27.69}{99.54}{123}{26.75}{100.05}{124}
\emline{26.75}{100.05}{125}{26.85}{100.53}{126}
\emline{26.85}{100.53}{127}{29.67}{101.33}{128}
\emline{31.67}{101.33}{129}{33.65}{100.90}{130}
\emline{33.65}{100.90}{131}{34.58}{100.44}{132}
\emline{34.58}{100.44}{133}{34.48}{99.95}{134}
\emline{34.48}{99.95}{135}{31.67}{99.00}{136}
\emline{29.67}{99.00}{137}{31.67}{99.00}{138}
\emline{29.67}{101.33}{139}{31.67}{101.33}{140}
\emline{29.67}{86.00}{141}{27.69}{86.54}{142}
\emline{27.69}{86.54}{143}{26.75}{87.05}{144}
\emline{26.75}{87.05}{145}{26.85}{87.53}{146}
\emline{26.85}{87.53}{147}{29.67}{88.33}{148}
\emline{31.67}{88.33}{149}{33.65}{87.90}{150}
\emline{33.65}{87.90}{151}{34.58}{87.44}{152}
\emline{34.58}{87.44}{153}{34.48}{86.95}{154}
\emline{34.48}{86.95}{155}{31.67}{86.00}{156}
\emline{29.67}{86.00}{157}{31.67}{86.00}{158}
\emline{29.67}{88.33}{159}{31.67}{88.33}{160}
\emline{27.00}{100.00}{161}{27.87}{97.74}{162}
\emline{27.87}{97.74}{163}{28.33}{95.48}{164}
\emline{28.33}{95.48}{165}{28.39}{93.21}{166}
\emline{28.39}{93.21}{167}{28.05}{90.95}{168}
\emline{28.05}{90.95}{169}{26.67}{87.33}{170}
\emline{34.67}{100.00}{171}{33.69}{97.74}{172}
\emline{33.69}{97.74}{173}{33.14}{95.48}{174}
\emline{33.14}{95.48}{175}{33.01}{93.21}{176}
\emline{33.01}{93.21}{177}{33.31}{90.95}{178}
\emline{33.31}{90.95}{179}{34.67}{87.33}{180}
\emline{63.00}{99.00}{181}{61.02}{99.54}{182}
\emline{61.02}{99.54}{183}{60.08}{100.05}{184}
\emline{60.08}{100.05}{185}{60.19}{100.53}{186}
\emline{60.19}{100.53}{187}{63.00}{101.33}{188}
\emline{65.00}{101.33}{189}{66.98}{100.90}{190}
\emline{66.98}{100.90}{191}{67.92}{100.44}{192}
\emline{67.92}{100.44}{193}{67.81}{99.95}{194}
\emline{67.81}{99.95}{195}{65.00}{99.00}{196}
\emline{63.00}{99.00}{197}{65.00}{99.00}{198}
\emline{63.00}{101.33}{199}{65.00}{101.33}{200}
\emline{63.00}{86.00}{201}{61.02}{86.54}{202}
\emline{61.02}{86.54}{203}{60.08}{87.05}{204}
\emline{60.08}{87.05}{205}{60.19}{87.53}{206}
\emline{60.19}{87.53}{207}{63.00}{88.33}{208}
\emline{65.00}{88.33}{209}{66.98}{87.90}{210}
\emline{66.98}{87.90}{211}{67.92}{87.44}{212}
\emline{67.92}{87.44}{213}{67.81}{86.95}{214}
\emline{67.81}{86.95}{215}{65.00}{86.00}{216}
\emline{63.00}{86.00}{217}{65.00}{86.00}{218}
\emline{63.00}{88.33}{219}{65.00}{88.33}{220}
\emline{60.33}{100.00}{221}{61.20}{97.74}{222}
\emline{61.20}{97.74}{223}{61.67}{95.48}{224}
\emline{61.67}{95.48}{225}{61.73}{93.21}{226}
\emline{61.73}{93.21}{227}{61.39}{90.95}{228}
\emline{61.39}{90.95}{229}{60.00}{87.33}{230}
\emline{68.00}{100.00}{231}{67.02}{97.74}{232}
\emline{67.02}{97.74}{233}{66.47}{95.48}{234}
\emline{66.47}{95.48}{235}{66.34}{93.21}{236}
\emline{66.34}{93.21}{237}{66.64}{90.95}{238}
\emline{66.64}{90.95}{239}{68.00}{87.33}{240}
\emline{71.33}{99.00}{241}{69.35}{99.54}{242}
\emline{69.35}{99.54}{243}{68.41}{100.05}{244}
\emline{68.41}{100.05}{245}{68.52}{100.53}{246}
\emline{68.52}{100.53}{247}{71.33}{101.33}{248}
\emline{73.33}{101.33}{249}{75.31}{100.90}{250}
\emline{75.31}{100.90}{251}{76.25}{100.44}{252}
\emline{76.25}{100.44}{253}{76.14}{99.95}{254}
\emline{76.14}{99.95}{255}{73.33}{99.00}{256}
\emline{71.33}{99.00}{257}{73.33}{99.00}{258}
\emline{71.33}{101.33}{259}{73.33}{101.33}{260}
\emline{71.33}{86.00}{261}{69.35}{86.54}{262}
\emline{69.35}{86.54}{263}{68.41}{87.05}{264}
\emline{68.41}{87.05}{265}{68.52}{87.53}{266}
\emline{68.52}{87.53}{267}{71.33}{88.33}{268}
\emline{73.33}{88.33}{269}{75.31}{87.90}{270}
\emline{75.31}{87.90}{271}{76.25}{87.44}{272}
\emline{76.25}{87.44}{273}{76.14}{86.95}{274}
\emline{76.14}{86.95}{275}{73.33}{86.00}{276}
\emline{71.33}{86.00}{277}{73.33}{86.00}{278}
\emline{71.33}{88.33}{279}{73.33}{88.33}{280}
\emline{68.66}{100.00}{281}{69.53}{97.74}{282}
\emline{69.53}{97.74}{283}{70.00}{95.48}{284}
\emline{70.00}{95.48}{285}{70.06}{93.21}{286}
\emline{70.06}{93.21}{287}{69.72}{90.95}{288}
\emline{69.72}{90.95}{289}{68.33}{87.33}{290}
\emline{76.33}{100.00}{291}{75.35}{97.74}{292}
\emline{75.35}{97.74}{293}{74.80}{95.48}{294}
\emline{74.80}{95.48}{295}{74.67}{93.21}{296}
\emline{74.67}{93.21}{297}{74.97}{90.95}{298}
\emline{74.97}{90.95}{299}{76.33}{87.33}{300}
\emline{79.33}{98.67}{301}{77.35}{99.21}{302}
\emline{77.35}{99.21}{303}{76.41}{99.72}{304}
\emline{76.41}{99.72}{305}{76.52}{100.20}{306}
\emline{76.52}{100.20}{307}{79.33}{101.00}{308}
\emline{81.33}{101.00}{309}{83.31}{100.57}{310}
\emline{83.31}{100.57}{311}{84.25}{100.11}{312}
\emline{84.25}{100.11}{313}{84.14}{99.62}{314}
\emline{84.14}{99.62}{315}{81.33}{98.67}{316}
\emline{79.33}{98.67}{317}{81.33}{98.67}{318}
\emline{79.33}{101.00}{319}{81.33}{101.00}{320}
\emline{79.33}{85.67}{321}{77.35}{86.21}{322}
\emline{77.35}{86.21}{323}{76.41}{86.72}{324}
\emline{76.41}{86.72}{325}{76.52}{87.20}{326}
\emline{76.52}{87.20}{327}{79.33}{88.00}{328}
\emline{81.33}{88.00}{329}{83.31}{87.57}{330}
\emline{83.31}{87.57}{331}{84.25}{87.11}{332}
\emline{84.25}{87.11}{333}{84.14}{86.62}{334}
\emline{84.14}{86.62}{335}{81.33}{85.67}{336}
\emline{79.33}{85.67}{337}{81.33}{85.67}{338}
\emline{79.33}{88.00}{339}{81.33}{88.00}{340}
\emline{76.66}{99.67}{341}{77.53}{97.41}{342}
\emline{77.53}{97.41}{343}{78.00}{95.15}{344}
\emline{78.00}{95.15}{345}{78.06}{92.88}{346}
\emline{78.06}{92.88}{347}{77.72}{90.62}{348}
\emline{77.72}{90.62}{349}{76.33}{87.00}{350}
\emline{84.33}{99.67}{351}{83.35}{97.41}{352}
\emline{83.35}{97.41}{353}{82.80}{95.15}{354}
\emline{82.80}{95.15}{355}{82.67}{92.88}{356}
\emline{82.67}{92.88}{357}{82.97}{90.62}{358}
\emline{82.97}{90.62}{359}{84.33}{87.00}{360}
\emline{5.34}{104.33}{361}{40.67}{104.33}{362}
\emline{34.34}{97.00}{363}{-1.00}{97.00}{364}
\emline{4.67}{91.00}{365}{40.00}{91.00}{366}
\emline{33.67}{83.67}{367}{-1.66}{83.67}{368}
\put(38.67,93.33){\circle*{0.67}}
\put(40.67,93.33){\circle*{0.67}}
\put(43.00,93.33){\circle*{0.67}}
\put(-2.33,93.33){\circle*{0.67}}
\put(-0.33,93.33){\circle*{0.67}}
\put(2.00,93.33){\circle*{0.67}}
\emline{34.00}{83.67}{369}{35.33}{86.00}{370}
\emline{35.67}{86.00}{371}{37.67}{87.67}{372}
\emline{37.67}{87.67}{373}{39.33}{89.33}{374}
\emline{40.33}{91.00}{375}{39.67}{89.33}{376}
\emline{34.67}{97.00}{377}{36.00}{99.33}{378}
\emline{36.33}{99.33}{379}{38.33}{101.00}{380}
\emline{38.33}{101.00}{381}{40.00}{102.67}{382}
\emline{41.00}{104.33}{383}{40.33}{102.67}{384}
\emline{-1.67}{83.67}{385}{-0.33}{86.00}{386}
\emline{0.00}{86.00}{387}{2.00}{87.67}{388}
\emline{2.00}{87.67}{389}{3.67}{89.33}{390}
\emline{4.67}{91.00}{391}{4.00}{89.33}{392}
\emline{-1.33}{97.00}{393}{0.00}{99.33}{394}
\emline{0.33}{99.33}{395}{2.33}{101.00}{396}
\emline{2.33}{101.00}{397}{4.00}{102.67}{398}
\emline{5.00}{104.33}{399}{4.33}{102.67}{400}
\emline{55.33}{97.00}{401}{57.67}{99.33}{402}
\emline{57.67}{99.33}{403}{58.67}{102.00}{404}
\emline{58.67}{102.00}{405}{61.33}{103.67}{406}
\emline{61.33}{103.67}{407}{89.67}{103.67}{408}
\emline{89.67}{103.67}{409}{87.33}{102.00}{410}
\emline{87.33}{102.00}{411}{87.33}{100.00}{412}
\emline{87.33}{100.00}{413}{86.00}{97.33}{414}
\emline{86.00}{97.33}{415}{55.67}{97.33}{416}
\emline{55.00}{83.67}{417}{57.33}{86.00}{418}
\emline{57.33}{86.00}{419}{58.33}{88.67}{420}
\emline{58.33}{88.67}{421}{61.00}{90.33}{422}
\emline{89.33}{90.33}{423}{87.00}{88.67}{424}
\emline{87.00}{88.67}{425}{87.00}{86.67}{426}
\emline{87.00}{86.67}{427}{85.67}{84.00}{428}
\emline{85.67}{84.00}{429}{55.33}{84.00}{430}
\emline{62.00}{90.00}{431}{63.67}{90.00}{432}
\emline{64.67}{90.00}{433}{66.00}{90.00}{434}
\emline{67.00}{90.00}{435}{69.33}{90.00}{436}
\emline{70.67}{90.00}{437}{72.00}{90.00}{438}
\emline{73.00}{90.00}{439}{74.33}{90.00}{440}
\emline{75.33}{90.00}{441}{77.67}{90.00}{442}
\emline{78.33}{90.00}{443}{80.00}{90.00}{444}
\emline{81.67}{90.00}{445}{82.67}{90.00}{446}
\emline{83.67}{90.00}{447}{89.33}{90.00}{448}
\put(54.67,93.33){\circle*{0.67}}
\put(56.67,93.33){\circle*{0.67}}
\put(59.00,93.33){\circle*{0.67}}
\put(85.33,93.33){\circle*{0.67}}
\put(87.33,93.33){\circle*{0.67}}
\put(89.66,93.33){\circle*{0.67}}
\put(15.33,78.00){\makebox(0,0)[cc]{(a)}}
\put(72.00,78.00){\makebox(0,0)[cc]{(b)}}
\emline{112.67}{100.00}{449}{114.89}{100.65}{450}
\emline{114.89}{100.65}{451}{118.00}{100.00}{452}
\emline{112.67}{100.00}{453}{114.89}{99.35}{454}
\emline{114.89}{99.35}{455}{118.00}{100.00}{456}
\emline{118.33}{100.00}{457}{120.56}{100.65}{458}
\emline{120.56}{100.65}{459}{123.67}{100.00}{460}
\emline{118.33}{100.00}{461}{120.56}{99.35}{462}
\emline{120.56}{99.35}{463}{123.67}{100.00}{464}
\emline{124.00}{100.00}{465}{126.22}{100.65}{466}
\emline{126.22}{100.65}{467}{129.33}{100.00}{468}
\emline{124.00}{100.00}{469}{126.22}{99.35}{470}
\emline{126.22}{99.35}{471}{129.33}{100.00}{472}
\emline{112.67}{87.67}{473}{114.89}{88.31}{474}
\emline{114.89}{88.31}{475}{118.00}{87.67}{476}
\emline{112.67}{87.67}{477}{114.89}{87.02}{478}
\emline{114.89}{87.02}{479}{118.00}{87.67}{480}
\emline{118.33}{87.67}{481}{120.56}{88.31}{482}
\emline{120.56}{88.31}{483}{123.67}{87.67}{484}
\emline{118.33}{87.67}{485}{120.56}{87.02}{486}
\emline{120.56}{87.02}{487}{123.67}{87.67}{488}
\emline{124.00}{87.67}{489}{126.22}{88.31}{490}
\emline{126.22}{88.31}{491}{129.33}{87.67}{492}
\emline{124.00}{87.67}{493}{126.22}{87.02}{494}
\emline{126.22}{87.02}{495}{129.33}{87.67}{496}
\emline{111.67}{97.67}{497}{133.33}{97.67}{498}
\emline{135.00}{102.67}{499}{114.33}{102.67}{500}
\emline{114.33}{102.67}{501}{113.33}{101.33}{502}
\emline{113.33}{101.33}{503}{113.33}{100.00}{504}
\emline{113.33}{100.00}{505}{112.00}{97.67}{506}
\emline{111.67}{85.00}{507}{133.33}{85.00}{508}
\emline{114.33}{90.00}{509}{113.33}{88.67}{510}
\emline{113.33}{88.67}{511}{113.33}{87.33}{512}
\emline{113.33}{87.33}{513}{112.00}{85.00}{514}
\emline{118.33}{95.67}{515}{117.36}{94.05}{516}
\emline{117.36}{94.05}{517}{118.33}{91.00}{518}
\emline{118.33}{95.67}{519}{119.60}{94.20}{520}
\emline{119.60}{94.20}{521}{118.67}{91.33}{522}
\emline{124.00}{95.67}{523}{123.03}{94.05}{524}
\emline{123.03}{94.05}{525}{124.00}{91.00}{526}
\emline{124.00}{95.67}{527}{125.27}{94.20}{528}
\emline{125.27}{94.20}{529}{124.33}{91.33}{530}
\emline{129.33}{95.67}{531}{128.36}{94.05}{532}
\emline{128.36}{94.05}{533}{129.33}{91.00}{534}
\emline{113.00}{96.00}{535}{114.27}{94.53}{536}
\emline{114.27}{94.53}{537}{113.33}{91.67}{538}
\emline{114.67}{90.00}{539}{116.00}{90.00}{540}
\emline{117.33}{90.00}{541}{119.33}{90.00}{542}
\emline{120.67}{90.00}{543}{122.67}{90.00}{544}
\emline{124.33}{90.00}{545}{125.67}{90.00}{546}
\emline{127.67}{90.00}{547}{129.33}{90.00}{548}
\emline{131.00}{90.00}{549}{132.67}{90.00}{550}
\emline{133.67}{90.00}{551}{135.33}{90.00}{552}
\put(105.33,93.33){\circle*{0.67}}
\put(107.33,93.33){\circle*{0.67}}
\put(109.66,93.33){\circle*{0.67}}
\put(138.00,93.33){\circle*{0.67}}
\put(140.00,93.33){\circle*{0.67}}
\put(142.33,93.33){\circle*{0.67}}
\put(122.00,78){\makebox(0,0)[cc]{(c)}}
\emline{135.67}{102.67}{553}{134.67}{101.33}{554}
\emline{134.67}{101.33}{555}{134.67}{100.00}{556}
\emline{134.67}{100.00}{557}{133.33}{97.67}{558}
\emline{135.33}{90.00}{559}{134.33}{88.67}{560}
\emline{134.33}{88.67}{561}{134.33}{87.33}{562}
\emline{134.33}{87.33}{563}{133.00}{85.00}{564}
\emline{129.33}{100.00}{565}{131.56}{100.65}{566}
\emline{131.56}{100.65}{567}{134.67}{100.00}{568}
\emline{129.33}{100.00}{569}{131.56}{99.35}{570}
\emline{131.56}{99.35}{571}{134.67}{100.00}{572}
\emline{129.00}{87.67}{573}{131.22}{88.31}{574}
\emline{131.22}{88.31}{575}{134.33}{87.67}{576}
\emline{129.00}{87.67}{577}{131.22}{87.02}{578}
\emline{131.22}{87.02}{579}{134.33}{87.67}{580}
\emline{129.67}{95.67}{581}{130.93}{94.20}{582}
\emline{130.93}{94.20}{583}{130.00}{91.33}{584}
\emline{134.33}{95.67}{585}{133.36}{94.05}{586}
\emline{133.36}{94.05}{587}{134.33}{91.00}{588}
\put(18.33,108.33){\makebox(0,0)[cc]{$T < T_{cr}$}}
\put(72.33,108.33){\makebox(0,0)[cc]{$T=T_{cr}$}}
\put(124.67,108.33){\makebox(0,0)[cc]{$T>T_{cr}$}}
\end{picture}

\vspace{-7.3cm}

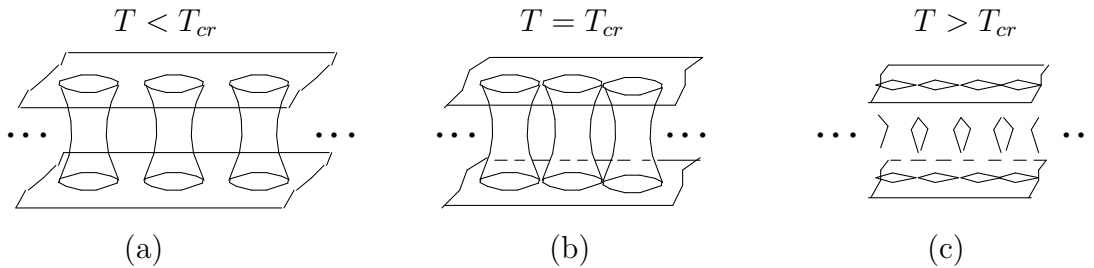
Fig.5: Pair production at finite temperature: (a) The region where the rate
exponent is independent of the temperature. (b) The critical point where the character
of the exponent behaviour changes. (c) The region where the temperature
dependence is essential.

\vspace{1.5cm}

The critical temperature, at which the $T$-dependence of the tunneling
rate drastically changes, can be easily found from the geometry:

\be
T_{cr} = \sqrt{1-E^2} \log\frac{1+E}{1-E}
\ee
It would be interesting to extract such critical temperature
from the partition function of the fundamental string
at nonzero temperature and the electric field found in \cite{tseytlin}.

\vspace{0.5cm}

\special{em:linewidth 0.4pt}
\unitlength 1.00mm
\linethickness{0.4pt}
\begin{picture}(96.33,108.33)
\emline{56.34}{99.00}{1}{54.36}{99.54}{2}
\emline{54.36}{99.54}{3}{53.42}{100.05}{4}
\emline{53.42}{100.05}{5}{53.53}{100.53}{6}
\emline{53.53}{100.53}{7}{56.34}{101.33}{8}
\emline{58.34}{101.33}{9}{60.32}{100.90}{10}
\emline{60.32}{100.90}{11}{61.26}{100.44}{12}
\emline{61.26}{100.44}{13}{61.15}{99.95}{14}
\emline{61.15}{99.95}{15}{58.34}{99.00}{16}
\emline{56.34}{99.00}{17}{58.34}{99.00}{18}
\emline{56.34}{101.33}{19}{58.34}{101.33}{20}
\emline{56.34}{86.00}{21}{54.36}{86.54}{22}
\emline{54.36}{86.54}{23}{53.42}{87.05}{24}
\emline{53.42}{87.05}{25}{53.53}{87.53}{26}
\emline{53.53}{87.53}{27}{56.34}{88.33}{28}
\emline{58.34}{88.33}{29}{60.32}{87.90}{30}
\emline{60.32}{87.90}{31}{61.26}{87.44}{32}
\emline{61.26}{87.44}{33}{61.15}{86.95}{34}
\emline{61.15}{86.95}{35}{58.34}{86.00}{36}
\emline{56.34}{86.00}{37}{58.34}{86.00}{38}
\emline{56.34}{88.33}{39}{58.34}{88.33}{40}
\emline{53.67}{99.67}{41}{54.54}{97.41}{42}
\emline{54.54}{97.41}{43}{55.00}{95.15}{44}
\emline{55.00}{95.15}{45}{55.06}{92.88}{46}
\emline{55.06}{92.88}{47}{54.72}{90.62}{48}
\emline{54.72}{90.62}{49}{53.33}{87.00}{50}
\emline{61.34}{100.00}{51}{60.36}{97.74}{52}
\emline{60.36}{97.74}{53}{59.81}{95.48}{54}
\emline{59.81}{95.48}{55}{59.68}{93.21}{56}
\emline{59.68}{93.21}{57}{59.98}{90.95}{58}
\emline{59.98}{90.95}{59}{61.34}{87.33}{60}
\emline{67.67}{99.00}{61}{65.69}{99.54}{62}
\emline{65.69}{99.54}{63}{64.75}{100.05}{64}
\emline{64.75}{100.05}{65}{64.86}{100.53}{66}
\emline{64.86}{100.53}{67}{67.67}{101.33}{68}
\emline{69.67}{101.33}{69}{71.65}{100.90}{70}
\emline{71.65}{100.90}{71}{72.59}{100.44}{72}
\emline{72.59}{100.44}{73}{72.48}{99.95}{74}
\emline{72.48}{99.95}{75}{69.67}{99.00}{76}
\emline{67.67}{99.00}{77}{69.67}{99.00}{78}
\emline{67.67}{101.33}{79}{69.67}{101.33}{80}
\emline{67.67}{86.00}{81}{65.69}{86.54}{82}
\emline{65.69}{86.54}{83}{64.75}{87.05}{84}
\emline{64.75}{87.05}{85}{64.86}{87.53}{86}
\emline{64.86}{87.53}{87}{67.67}{88.33}{88}
\emline{69.67}{88.33}{89}{71.65}{87.90}{90}
\emline{71.65}{87.90}{91}{72.59}{87.44}{92}
\emline{72.59}{87.44}{93}{72.48}{86.95}{94}
\emline{72.48}{86.95}{95}{69.67}{86.00}{96}
\emline{67.67}{86.00}{97}{69.67}{86.00}{98}
\emline{67.67}{88.33}{99}{69.67}{88.33}{100}
\emline{65.00}{100.00}{101}{65.87}{97.74}{102}
\emline{65.87}{97.74}{103}{66.34}{95.48}{104}
\emline{66.34}{95.48}{105}{66.40}{93.21}{106}
\emline{66.40}{93.21}{107}{66.06}{90.95}{108}
\emline{66.06}{90.95}{109}{64.67}{87.33}{110}
\emline{72.67}{100.00}{111}{71.69}{97.74}{112}
\emline{71.69}{97.74}{113}{71.14}{95.48}{114}
\emline{71.14}{95.48}{115}{71.01}{93.21}{116}
\emline{71.01}{93.21}{117}{71.31}{90.95}{118}
\emline{71.31}{90.95}{119}{72.67}{87.33}{120}
\emline{79.00}{99.00}{121}{77.02}{99.54}{122}
\emline{77.02}{99.54}{123}{76.08}{100.05}{124}
\emline{76.08}{100.05}{125}{76.19}{100.53}{126}
\emline{76.19}{100.53}{127}{79.00}{101.33}{128}
\emline{81.00}{101.33}{129}{82.98}{100.90}{130}
\emline{82.98}{100.90}{131}{83.92}{100.44}{132}
\emline{83.92}{100.44}{133}{83.81}{99.95}{134}
\emline{83.81}{99.95}{135}{81.00}{99.00}{136}
\emline{79.00}{99.00}{137}{81.00}{99.00}{138}
\emline{79.00}{101.33}{139}{81.00}{101.33}{140}
\emline{79.00}{86.00}{141}{77.02}{86.54}{142}
\emline{77.02}{86.54}{143}{76.08}{87.05}{144}
\emline{76.08}{87.05}{145}{76.19}{87.53}{146}
\emline{76.19}{87.53}{147}{79.00}{88.33}{148}
\emline{81.00}{88.33}{149}{82.98}{87.90}{150}
\emline{82.98}{87.90}{151}{83.92}{87.44}{152}
\emline{83.92}{87.44}{153}{83.81}{86.95}{154}
\emline{83.81}{86.95}{155}{81.00}{86.00}{156}
\emline{79.00}{86.00}{157}{81.00}{86.00}{158}
\emline{79.00}{88.33}{159}{81.00}{88.33}{160}
\emline{76.33}{100.00}{161}{77.20}{97.74}{162}
\emline{77.20}{97.74}{163}{77.67}{95.48}{164}
\emline{77.67}{95.48}{165}{77.73}{93.21}{166}
\emline{77.73}{93.21}{167}{77.39}{90.95}{168}
\emline{77.39}{90.95}{169}{76.00}{87.33}{170}
\emline{84.00}{100.00}{171}{83.02}{97.74}{172}
\emline{83.02}{97.74}{173}{82.47}{95.48}{174}
\emline{82.47}{95.48}{175}{82.34}{93.21}{176}
\emline{82.34}{93.21}{177}{82.64}{90.95}{178}
\emline{82.64}{90.95}{179}{84.00}{87.33}{180}
\emline{54.67}{104.33}{181}{90.00}{104.33}{182}
\emline{83.67}{97.00}{183}{48.33}{97.00}{184}
\emline{84.00}{97.00}{185}{86.33}{99.00}{186}
\emline{87.33}{100.00}{187}{89.33}{102.00}{188}
\emline{90.00}{103.00}{189}{91.00}{104.33}{190}
\emline{48.67}{97.67}{191}{50.33}{99.33}{192}
\emline{51.33}{100.67}{193}{53.33}{103.00}{194}
\emline{83.00}{83.67}{195}{47.67}{83.67}{196}
\emline{83.33}{83.67}{197}{85.67}{85.67}{198}
\emline{86.67}{86.67}{199}{88.67}{88.67}{200}
\emline{89.33}{89.67}{201}{90.33}{91.00}{202}
\emline{48.00}{84.33}{203}{49.67}{86.00}{204}
\emline{50.67}{87.33}{205}{52.67}{89.67}{206}
\put(91.67,93.33){\circle*{0.67}}
\put(93.67,93.33){\circle*{0.67}}
\put(96.00,93.33){\circle*{0.67}}
\put(43.33,93.33){\circle*{0.67}}
\put(45.33,93.33){\circle*{0.67}}
\put(47.66,93.33){\circle*{0.67}}
\emline{89.00}{91.00}{207}{86.67}{91.00}{208}
\emline{85.00}{91.00}{209}{83.33}{91.00}{210}
\emline{80.67}{91.00}{211}{79.00}{91.00}{212}
\emline{77.00}{91.00}{213}{74.67}{91.00}{214}
\emline{73.00}{91.00}{215}{70.33}{91.00}{216}
\emline{68.67}{91.00}{217}{67.00}{91.00}{218}
\emline{65.67}{91.00}{219}{62.33}{91.00}{220}
\emline{60.67}{91.00}{221}{58.33}{91.00}{222}
\emline{56.67}{91.00}{223}{54.33}{91.00}{224}
\special{em:linewidth 1.4pt}
\emline{84.00}{87.00}{225}{87.33}{87.00}{226}
\emline{87.33}{87.00}{227}{87.33}{100.33}{228}
\emline{87.33}{100.33}{229}{84.00}{100.33}{230}
\emline{76.00}{100.33}{231}{72.67}{100.33}{232}
\emline{64.67}{100.33}{233}{61.33}{100.33}{234}
\emline{53.00}{100.33}{235}{51.33}{100.33}{236}
\emline{51.33}{100.33}{237}{51.00}{87.33}{238}
\emline{51.00}{87.33}{239}{53.33}{87.33}{240}
\emline{61.33}{87.33}{241}{64.67}{87.00}{242}
\emline{72.67}{87.00}{243}{76.33}{87.00}{244}
\end{picture}

\vspace{-7.8cm}
\noindent
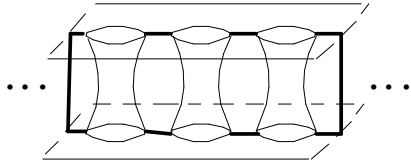
Fig.6: Pair creation with nonvanishing total magnetic or electric charges  at finite temperature.

\vspace{1cm}

Another interesting  process at finite temperature is
creation of the pair with nonvanishing total
magnetic or electric charges. The simplest geometry
of such process which takes use of junctions looks as follows.
The array now consists of the deformed cylinders formed from
the $(p_1,q_1)$ and $(p_2,q_2)$-strings while at small
temperatures they are connected by the sheets of
$(p_1+p_2,q_1+q_2)$-strings (Fig.6). This configuration
implies the nontrivial temperature dependence
for the corresponding amplitude at any temperature.

\vspace{0.2cm}

\section{Decay of BPS particles in the external fields}

In this section we consider novel phenomena
of the BPS particles decay at rest in the external
fields. We concentrate on the examples of the
monopole decay in the electric field and the charge decay in
the magnetic field. However the consideration of the
generic BPS particle decay can be treated in the similar
manner\footnote{We call these precesses "decay" although a different interpretation
is possible. Monopole decay can be considered as a transition of the
monopole into dyon and charge, or as an induced emission of the charge
by the monopole. Charge decay can be considered as induced production
of the dyon-antimonopole pair.}.
Note that the case considered here differs from the
process of the monopole decay in the external
$B_{\mu\nu}$ field in the bulk with the constant curvature
considered in \cite{gs1}.

Let us explain the geometry of the process. We have
BPS monopole at rest represented by the  D1 string stretched
between the two D3 branes.
The electric field is switched on  both D3 branes but evidently D1
brane classically
doesn't feel it.  However D1 string can split into F1 string
and $(-1,1)$ string due to junction. After some interval
of the Euclidean time virtual strings join into D1
again and the whole Euclidean solution looks like surface
with the  topology of the cylinder between two semiplanes corresponding
to the worldvolumes of D1 strings. Brane configurations
corresponding to the monopole and charge decays are presented at Fig.7a-b.

\vspace{1cm}

\special{em:linewidth 0.4pt}
\unitlength 1.00mm
\linethickness{0.4pt}
\begin{picture}(143.00,94.67)
\emline{5.67}{80.00}{1}{21.33}{89.00}{2}
\emline{21.33}{89.00}{3}{61.33}{89.00}{4}
\emline{61.33}{89.00}{5}{46.33}{80.33}{6}
\emline{46.33}{80.33}{7}{6.00}{80.33}{8}
\emline{27.67}{86.00}{9}{30.19}{86.41}{10}
\emline{30.19}{86.41}{11}{32.65}{86.63}{12}
\emline{32.65}{86.63}{13}{35.03}{86.65}{14}
\emline{35.03}{86.65}{15}{37.33}{86.47}{16}
\emline{37.33}{86.47}{17}{40.00}{86.00}{18}
\emline{28.00}{86.00}{19}{29.35}{84.24}{20}
\emline{29.35}{84.24}{21}{30.75}{82.98}{22}
\emline{30.75}{82.98}{23}{32.19}{82.20}{24}
\emline{32.19}{82.20}{25}{33.67}{81.92}{26}
\emline{33.67}{81.92}{27}{35.19}{82.12}{28}
\emline{35.19}{82.12}{29}{36.75}{82.81}{30}
\emline{36.75}{82.81}{31}{38.35}{84.00}{32}
\emline{38.35}{84.00}{33}{40.00}{85.67}{34}
\emline{61.33}{70.00}{35}{46.33}{61.33}{36}
\emline{46.33}{61.33}{37}{6.00}{61.33}{38}
\emline{27.67}{67.00}{39}{30.19}{67.41}{40}
\emline{30.19}{67.41}{41}{32.65}{67.63}{42}
\emline{32.65}{67.63}{43}{35.03}{67.65}{44}
\emline{35.03}{67.65}{45}{37.33}{67.47}{46}
\emline{37.33}{67.47}{47}{40.00}{67.00}{48}
\emline{28.00}{67.00}{49}{29.35}{65.24}{50}
\emline{29.35}{65.24}{51}{30.75}{63.98}{52}
\emline{30.75}{63.98}{53}{32.19}{63.20}{54}
\emline{32.19}{63.20}{55}{33.67}{62.92}{56}
\emline{33.67}{62.92}{57}{35.19}{63.12}{58}
\emline{35.19}{63.12}{59}{36.75}{63.81}{60}
\emline{36.75}{63.81}{61}{38.35}{65.00}{62}
\emline{38.35}{65.00}{63}{40.00}{66.67}{64}
\emline{40.33}{85.67}{65}{39.38}{83.34}{66}
\emline{39.38}{83.34}{67}{38.69}{81.00}{68}
\emline{38.69}{81.00}{69}{38.25}{78.67}{70}
\emline{38.25}{78.67}{71}{38.08}{76.33}{72}
\emline{38.08}{76.33}{73}{38.17}{74.00}{74}
\emline{38.17}{74.00}{75}{38.52}{71.67}{76}
\emline{38.52}{71.67}{77}{39.13}{69.33}{78}
\emline{39.13}{69.33}{79}{40.00}{67.00}{80}
\emline{27.67}{85.67}{81}{28.51}{83.44}{82}
\emline{28.51}{83.44}{83}{29.12}{81.20}{84}
\emline{29.12}{81.20}{85}{29.51}{78.96}{86}
\emline{29.51}{78.96}{87}{29.67}{76.70}{88}
\emline{29.67}{76.70}{89}{29.60}{74.44}{90}
\emline{29.60}{74.44}{91}{29.30}{72.16}{92}
\emline{29.30}{72.16}{93}{28.78}{69.88}{94}
\emline{28.78}{69.88}{95}{27.67}{66.67}{96}
\emline{6.00}{61.33}{97}{15.33}{66.67}{98}
\emline{16.67}{67.33}{99}{18.67}{68.33}{100}
\emline{19.33}{69.00}{101}{21.33}{70.00}{102}
\emline{21.33}{70.00}{103}{23.33}{70.00}{104}
\emline{25.00}{70.00}{105}{27.67}{70.00}{106}
\emline{28.67}{70.00}{107}{30.33}{70.00}{108}
\emline{32.33}{70.00}{109}{35.00}{70.00}{110}
\emline{36.33}{70.00}{111}{38.33}{70.00}{112}
\emline{40.33}{70.00}{113}{43.33}{70.00}{114}
\emline{46.33}{70.00}{115}{49.33}{70.00}{116}
\emline{52.00}{70.00}{117}{55.33}{70.00}{118}
\emline{61.33}{70.00}{119}{56.33}{70.00}{120}
\emline{84.67}{80.00}{121}{100.33}{89.00}{122}
\emline{100.33}{89.00}{123}{140.33}{89.00}{124}
\emline{140.33}{89.00}{125}{125.33}{80.33}{126}
\emline{125.33}{80.33}{127}{85.00}{80.33}{128}
\emline{140.33}{70.00}{129}{125.33}{61.33}{130}
\emline{125.33}{61.33}{131}{85.00}{61.33}{132}
\emline{119.33}{84.67}{133}{118.38}{82.34}{134}
\emline{118.38}{82.34}{135}{117.69}{80.00}{136}
\emline{117.69}{80.00}{137}{117.25}{77.67}{138}
\emline{117.25}{77.67}{139}{117.08}{75.33}{140}
\emline{117.08}{75.33}{141}{117.17}{73.00}{142}
\emline{117.17}{73.00}{143}{117.52}{70.67}{144}
\emline{117.52}{70.67}{145}{118.13}{68.33}{146}
\emline{118.13}{68.33}{147}{119.00}{66.00}{148}
\emline{106.33}{84.67}{149}{107.17}{82.44}{150}
\emline{107.17}{82.44}{151}{107.78}{80.20}{152}
\emline{107.78}{80.20}{153}{108.17}{77.96}{154}
\emline{108.17}{77.96}{155}{108.33}{75.70}{156}
\emline{108.33}{75.70}{157}{108.26}{73.44}{158}
\emline{108.26}{73.44}{159}{107.96}{71.16}{160}
\emline{107.96}{71.16}{161}{107.44}{68.88}{162}
\emline{107.44}{68.88}{163}{106.33}{65.67}{164}
\emline{85.00}{61.33}{165}{94.33}{66.67}{166}
\emline{95.67}{67.33}{167}{97.67}{68.33}{168}
\emline{98.33}{69.00}{169}{100.33}{70.00}{170}
\emline{100.33}{70.00}{171}{102.33}{70.00}{172}
\emline{104.00}{70.00}{173}{106.67}{70.00}{174}
\emline{107.67}{70.00}{175}{109.33}{70.00}{176}
\emline{111.33}{70.00}{177}{114.00}{70.00}{178}
\emline{115.33}{70.00}{179}{117.33}{70.00}{180}
\emline{119.33}{70.00}{181}{122.33}{70.00}{182}
\emline{125.33}{70.00}{183}{128.33}{70.00}{184}
\emline{131.00}{70.00}{185}{134.33}{70.00}{186}
\emline{140.33}{70.00}{187}{135.33}{70.00}{188}
\emline{106.67}{84.67}{189}{108.78}{85.78}{190}
\emline{108.78}{85.78}{191}{110.89}{86.45}{192}
\emline{110.89}{86.45}{193}{115.11}{86.45}{194}
\emline{115.11}{86.45}{195}{117.22}{85.78}{196}
\emline{117.22}{85.78}{197}{119.33}{84.67}{198}
\emline{106.67}{84.33}{199}{108.82}{83.35}{200}
\emline{108.82}{83.35}{201}{111.00}{82.80}{202}
\emline{111.00}{82.80}{203}{113.19}{82.67}{204}
\emline{113.19}{82.67}{205}{115.41}{82.97}{206}
\emline{115.41}{82.97}{207}{119.00}{84.33}{208}
\emline{106.33}{66.00}{209}{108.44}{67.11}{210}
\emline{108.44}{67.11}{211}{110.56}{67.78}{212}
\emline{110.56}{67.78}{213}{114.78}{67.78}{214}
\emline{114.78}{67.78}{215}{116.89}{67.11}{216}
\emline{116.89}{67.11}{217}{119.00}{66.00}{218}
\emline{106.33}{65.67}{219}{108.48}{64.69}{220}
\emline{108.48}{64.69}{221}{110.66}{64.14}{222}
\emline{110.66}{64.14}{223}{112.86}{64.01}{224}
\emline{112.86}{64.01}{225}{115.07}{64.31}{226}
\emline{115.07}{64.31}{227}{118.67}{65.67}{228}
\special{em:linewidth 1.2pt}
\emline{15.67}{85.67}{229}{27.67}{85.67}{230}
\emline{40.00}{85.67}{231}{55.33}{85.67}{232}
\emline{55.33}{85.67}{233}{55.33}{66.33}{234}
\emline{55.33}{66.33}{235}{40.33}{66.33}{236}
\emline{27.67}{66.33}{237}{15.00}{66.33}{238}
\emline{15.00}{66.33}{239}{15.00}{85.67}{240}
\emline{106.33}{65.67}{241}{92.67}{65.67}{242}
\emline{92.67}{65.67}{243}{92.67}{84.67}{244}
\emline{92.67}{84.67}{245}{106.67}{84.67}{246}
\emline{119.00}{84.67}{247}{133.33}{84.67}{248}
\emline{133.33}{84.67}{249}{133.33}{66.00}{250}
\emline{133.33}{66.00}{251}{119.00}{66.00}{252}
\special{em:linewidth 0.4pt}
\put(100.00,74.67){\makebox(0,0)[cc]{\footnotesize $(1,0)$}}
\emline{114.33}{86.67}{253}{114.29}{88.59}{254}
\emline{114.29}{88.59}{255}{113.74}{90.11}{256}
\emline{113.74}{90.11}{257}{112.67}{91.25}{258}
\emline{112.67}{91.25}{259}{111.07}{92.00}{260}
\emline{111.07}{92.00}{261}{108.96}{92.36}{262}
\emline{108.96}{92.36}{263}{106.33}{92.33}{264}
\put(106.33,94.67){\makebox(0,0)[cc]{\footnotesize $(1,1)$}}
\emline{111.33}{64.00}{265}{111.60}{62.29}{266}
\emline{111.60}{62.29}{267}{112.25}{60.81}{268}
\emline{112.25}{60.81}{269}{113.27}{59.58}{270}
\emline{113.27}{59.58}{271}{114.67}{58.59}{272}
\emline{114.67}{58.59}{273}{116.44}{57.83}{274}
\emline{116.44}{57.83}{275}{118.58}{57.31}{276}
\emline{118.58}{57.31}{277}{121.10}{57.04}{278}
\emline{121.10}{57.04}{279}{124.00}{57.00}{280}
\put(129.67,58.00){\makebox(0,0)[cc]{\footnotesize $(0,-1)$}}
\put(110.67,48.00){\makebox(0,0)[cc]{(b)}}
\put(33.67,48.00){\makebox(0,0)[cc]{(a)}}
\emline{36.33}{63.67}{281}{36.60}{61.95}{282}
\emline{36.60}{61.95}{283}{37.25}{60.48}{284}
\emline{37.25}{60.48}{285}{38.27}{59.25}{286}
\emline{38.27}{59.25}{287}{39.67}{58.25}{288}
\emline{39.67}{58.25}{289}{41.44}{57.50}{290}
\emline{41.44}{57.50}{291}{43.58}{56.98}{292}
\emline{43.58}{56.98}{293}{46.10}{56.70}{294}
\emline{46.10}{56.70}{295}{49.00}{56.67}{296}
\put(0.00,0.00){}
\put(54.00,56.67){\makebox(0,0)[cc]{\footnotesize $(1,0)$}}
\emline{33.00}{86.67}{297}{32.96}{88.59}{298}
\emline{32.96}{88.59}{299}{32.41}{90.11}{300}
\emline{32.41}{90.11}{301}{31.34}{91.25}{302}
\emline{31.34}{91.25}{303}{29.74}{92.00}{304}
\emline{29.74}{92.00}{305}{27.63}{92.36}{306}
\emline{27.63}{92.36}{307}{25.00}{92.33}{308}
\put(25.00,94.67){\makebox(0,0)[cc]{\footnotesize $(-1,1)$}}
\put(21.67,75.00){\makebox(0,0)[cc]{\footnotesize $(0,1)$}}
\put(145.00,90.33){\makebox(0,0)[cc]{D3}}
\put(145.0,69.67){\makebox(0,0)[cc]{D3}}
\put(65.00,90.33){\makebox(0,0)[cc]{D3}}
\put(65.0,69.67){\makebox(0,0)[cc]{D3}}
\end{picture}

\vspace{-4.6cm}

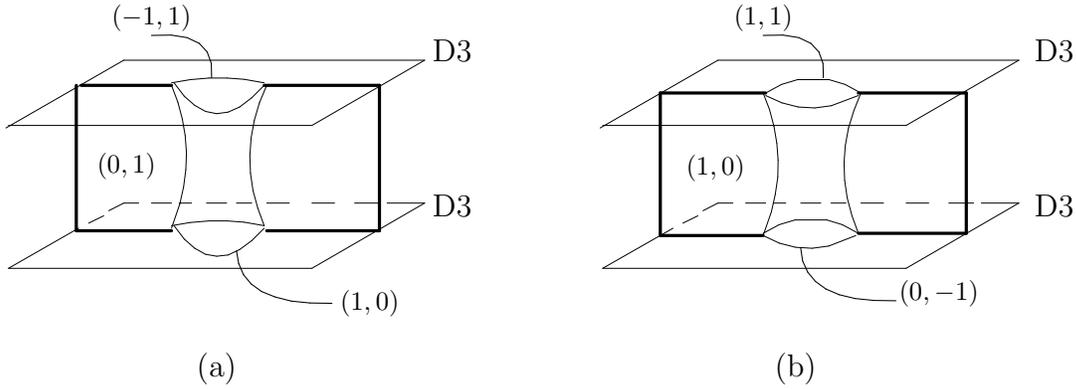
Fig.7: Induced decay of  BPS particles in the external fields in the brane
language:\\
(a) Monopole $(0,1)$ transition into dyon $(-1,1)$ and charge $(1,0)$ in the constant
electric field.
(b) Electric charge $(1,0)$ decay into dyon $(1,1)$ and anti-monopole $(0,-1)$ in the
constant magnetic field.

\vspace{1cm}

Let us turn now to calculation of the probabilities of such processes
with the exponential accuracy. We start with the monopole decay
in the weak electric field. At the weak coupling the tension
of the D1 and dyonic string is much larger then the tension
of F1 string. Therefore F1 string joins at junction  almost at the
right angle. Hence in the weak field worldvolume of the virtual
F1 string looks as a half of the cylinder while the dyonic string
worldvolume can be considered as plane. The sum of the both
contributions amounts into the following probability

\be
w_{mon} \propto \exp \left( 2\pi R v (T_{0,1}-T_{-1,1}) - \pi R v T_{1,0}\right)
\ee
where $R$ is the radius of the Euclidean trajectory of the
charge. Since $\delta T $ is proportional to the coupling
constant, the first term can be neglected and the
probability rate equals to the square root of the spontaneous
charge production in the electric field

\be
w_{mon}=\sqrt{ w_{spon}}
\ee
To get the probability of the charge decay in the magnetic
field note that now both  virtual strings are heavy
at the weak coupling and the
Euclidian configuration of the virtual strings almost coincides
with the configuration corresponding  to the spontaneous
production of the monopole pair in the magnetic field. Therefore
the probabilities of these two processes coincide as well.
Let us emphasize that the probability rates for both processes can be found
in the purely field theory approach without any
reference to the string theory. The field theory
counterpart of the junction is the process
when the charged state  fills the bound state on the
monopole.

One more interesting process
in the magnetic field can be considered when
the fundamental matter is added. Let us consider the
quark in the initial state represented by the string
stretched between D3 and D7 brane. The junction
amounts to the virtual D1 and dyonic strings forming a
cylinder surface. This physical process corresponds to
the decay of the quark into the magnetically charged
states in the background field.

Let us also
suggest the qualitative
field theory explanation of the factor $1/2$ in the
exponent of the monopole decay probability. To this aim
remind that there are discrete levels
of the charge in the monopole
background just at the middle of the forbidden zone. Therefore, by
taking into account the fact that the charge yields the main contribution
to the action, we immediately arrive at the desired factor
since the Euclidean path from the discrete level is twice
smaller then the path from the lower branch of the Dirac sea.

It is interesting to discuss decay processes at almost
critical  fields. Since the charge decay process
is described by the almost radially symmetric brane
configuration it is clear that at the critical
magnetic field the surface shrinks at one point and
two D1 strings seem to be asymptotic states of the decay.
In the monopole decay case the picture is more subtle.
Indeed since there is no radial symmetry it is
difficult to get the explicit form of the minimal surface.
However it seems that the critical electric field such that the
shrinking occurs, does exist. The point is that the
main contribution to the action comes from the charge
and shrinking of the "half" of the surface should
amounts into the shrinking of the whole surface.


\section{Pair production in  gauge theories at large $N$
 and \\ AdS/CFT correspondence}

\subsection{Set-up}

In this section we study W-boson pair creation in \N4 SUSYM
theory in strong coupling regime using the AdS/CFT correspondence~\cite{mald}.

\vspace{1cm}

\special{em:linewidth 1.4pt}
\unitlength 1.00mm
\linethickness{0.4pt}
\begin{picture}(123.33,92.00)
\special{em:linewidth 1.4pt}
\emline{10.00}{50.00}{1}{10.00}{80.00}{2}
\emline{10.00}{80.00}{3}{20.00}{90.00}{4}
\emline{20.00}{60.00}{5}{10.00}{50.00}{6}
\special{em:linewidth 0.4pt}
\emline{30.00}{50.00}{7}{30.00}{80.00}{8}
\emline{30.00}{80.00}{9}{40.00}{90.00}{10}
\emline{40.00}{90.00}{11}{40.00}{60.00}{12}
\emline{40.00}{60.00}{13}{30.00}{50.00}{14}
\put(20.00,90.00){\vector(-1,0){0.2}}
\emline{30.00}{90.00}{15}{20.00}{90.00}{16}
\put(40.00,90.00){\vector(1,0){0.2}}
\emline{29.67}{90.00}{17}{40.00}{90.00}{18}
\put(28.67,92.00){\makebox(0,0)[cc]{$v$}}
\emline{34.00}{70.67}{19}{34.20}{73.09}{20}
\emline{34.20}{73.09}{21}{34.40}{75.14}{22}
\emline{34.40}{75.14}{23}{34.60}{76.82}{24}
\emline{34.60}{76.82}{25}{34.81}{78.15}{26}
\emline{34.81}{78.15}{27}{35.02}{79.11}{28}
\emline{35.02}{79.11}{29}{35.24}{79.70}{30}
\emline{35.24}{79.70}{31}{35.46}{79.93}{32}
\emline{35.46}{79.93}{33}{35.68}{79.80}{34}
\emline{35.68}{79.80}{35}{35.91}{79.30}{36}
\emline{35.91}{79.30}{37}{36.14}{78.44}{38}
\emline{36.14}{78.44}{39}{36.37}{77.22}{40}
\emline{36.37}{77.22}{41}{36.61}{75.63}{42}
\emline{36.61}{75.63}{43}{37.00}{72.33}{44}
\emline{34.00}{67.33}{45}{34.42}{65.16}{46}
\emline{34.42}{65.16}{47}{34.80}{63.40}{48}
\emline{34.80}{63.40}{49}{35.15}{62.03}{50}
\emline{35.15}{62.03}{51}{35.48}{61.06}{52}
\emline{35.48}{61.06}{53}{35.77}{60.49}{54}
\emline{35.77}{60.49}{55}{36.04}{60.32}{56}
\emline{36.04}{60.32}{57}{36.27}{60.55}{58}
\emline{36.27}{60.55}{59}{36.47}{61.18}{60}
\emline{36.47}{61.18}{61}{36.64}{62.20}{62}
\emline{36.64}{62.20}{63}{36.78}{63.63}{64}
\emline{36.78}{63.63}{65}{36.90}{65.45}{66}
\emline{36.90}{65.45}{67}{37.00}{68.67}{68}
\emline{34.00}{70.33}{69}{34.00}{67.33}{70}
\emline{37.00}{72.33}{71}{37.00}{66.67}{72}
\put(31.67,44.67){\makebox(0,0)[cc]{D3}}
\put(9.67,44.33){\makebox(0,0)[cc]{$N$ D3}}
\emline{13.33}{70.67}{73}{13.53}{73.09}{74}
\emline{13.53}{73.09}{75}{13.73}{75.14}{76}
\emline{13.73}{75.14}{77}{13.94}{76.82}{78}
\emline{13.94}{76.82}{79}{14.15}{78.15}{80}
\emline{14.15}{78.15}{81}{14.36}{79.11}{82}
\emline{14.36}{79.11}{83}{14.57}{79.70}{84}
\emline{14.57}{79.70}{85}{14.79}{79.93}{86}
\emline{14.79}{79.93}{87}{15.02}{79.80}{88}
\emline{15.02}{79.80}{89}{15.24}{79.30}{90}
\emline{15.24}{79.30}{91}{15.47}{78.44}{92}
\emline{15.47}{78.44}{93}{15.71}{77.22}{94}
\emline{15.71}{77.22}{95}{15.94}{75.63}{96}
\emline{15.94}{75.63}{97}{16.33}{72.33}{98}
\emline{13.33}{67.33}{99}{13.75}{65.16}{100}
\emline{13.75}{65.16}{101}{14.13}{63.40}{102}
\emline{14.13}{63.40}{103}{14.48}{62.03}{104}
\emline{14.48}{62.03}{105}{14.81}{61.06}{106}
\emline{14.81}{61.06}{107}{15.10}{60.49}{108}
\emline{15.10}{60.49}{109}{15.37}{60.32}{110}
\emline{15.37}{60.32}{111}{15.60}{60.55}{112}
\emline{15.60}{60.55}{113}{15.80}{61.18}{114}
\emline{15.80}{61.18}{115}{15.97}{62.20}{116}
\emline{15.97}{62.20}{117}{16.11}{63.63}{118}
\emline{16.11}{63.63}{119}{16.23}{65.45}{120}
\emline{16.23}{65.45}{121}{16.33}{68.67}{122}
\emline{13.33}{70.33}{123}{13.33}{67.33}{124}
\emline{16.33}{72.33}{125}{16.33}{66.67}{126}
\emline{15.00}{80.00}{127}{16.95}{78.80}{128}
\emline{16.95}{78.80}{129}{18.92}{77.87}{130}
\emline{18.92}{77.87}{131}{20.92}{77.20}{132}
\emline{20.92}{77.20}{133}{22.95}{76.80}{134}
\emline{22.95}{76.80}{135}{27.08}{76.80}{136}
\emline{27.08}{76.80}{137}{29.19}{77.20}{138}
\emline{29.19}{77.20}{139}{31.32}{77.87}{140}
\emline{31.32}{77.87}{141}{33.48}{78.80}{142}
\emline{33.48}{78.80}{143}{35.67}{80.00}{144}
\emline{15.33}{60.00}{145}{17.02}{61.46}{146}
\emline{17.02}{61.46}{147}{18.75}{62.62}{148}
\emline{18.75}{62.62}{149}{20.54}{63.48}{150}
\emline{20.54}{63.48}{151}{22.36}{64.04}{152}
\emline{22.36}{64.04}{153}{24.24}{64.31}{154}
\emline{24.24}{64.31}{155}{26.15}{64.28}{156}
\emline{26.15}{64.28}{157}{28.12}{63.95}{158}
\emline{28.12}{63.95}{159}{30.13}{63.33}{160}
\emline{30.13}{63.33}{161}{32.18}{62.41}{162}
\emline{32.18}{62.41}{163}{36.00}{60.00}{164}
\put(24.33,69.67){\makebox(0,0)[cc]{F1}}
\special{em:linewidth 1.4pt}
\emline{20.00}{90.00}{165}{20.00}{77.67}{166}
\emline{20.00}{75.67}{167}{20.00}{73.33}{168}
\emline{20.00}{71.00}{169}{20.00}{69.00}{170}
\emline{20.00}{66.67}{171}{20.00}{65.00}{172}
\emline{20.00}{63.33}{173}{20.00}{60.33}{174}
\special{em:linewidth 0.4pt}
\put(78.00,70.00){\vector(1,0){0.2}}
\emline{54.00}{70.00}{175}{78.00}{70.00}{176}
\put(66.33,72.50){\makebox(0,0)[cc]{AdS/CFT}}
\put(22.67,30.00){\makebox(0,0)[cc]{(a)}}
\emline{113.33}{50.00}{177}{113.33}{80.00}{178}
\emline{113.33}{80.00}{179}{123.33}{90.00}{180}
\emline{123.33}{90.00}{181}{123.33}{60.00}{182}
\emline{123.33}{60.00}{183}{113.33}{50.00}{184}
\emline{117.33}{70.67}{185}{117.53}{73.09}{186}
\emline{117.53}{73.09}{187}{117.73}{75.14}{188}
\emline{117.73}{75.14}{189}{117.94}{76.82}{190}
\emline{117.94}{76.82}{191}{118.15}{78.15}{192}
\emline{118.15}{78.15}{193}{118.36}{79.11}{194}
\emline{118.36}{79.11}{195}{118.57}{79.70}{196}
\emline{118.57}{79.70}{197}{118.79}{79.93}{198}
\emline{118.79}{79.93}{199}{119.02}{79.80}{200}
\emline{119.02}{79.80}{201}{119.24}{79.30}{202}
\emline{119.24}{79.30}{203}{119.47}{78.44}{204}
\emline{119.47}{78.44}{205}{119.71}{77.22}{206}
\emline{119.71}{77.22}{207}{119.95}{75.63}{208}
\emline{119.95}{75.63}{209}{120.33}{72.33}{210}
\emline{117.33}{67.33}{211}{117.75}{65.16}{212}
\emline{117.75}{65.16}{213}{118.13}{63.40}{214}
\emline{118.13}{63.40}{215}{118.49}{62.03}{216}
\emline{118.49}{62.03}{217}{118.81}{61.06}{218}
\emline{118.81}{61.06}{219}{119.11}{60.49}{220}
\emline{119.11}{60.49}{221}{119.37}{60.32}{222}
\emline{119.37}{60.32}{223}{119.60}{60.55}{224}
\emline{119.60}{60.55}{225}{119.80}{61.18}{226}
\emline{119.80}{61.18}{227}{119.98}{62.20}{228}
\emline{119.98}{62.20}{229}{120.12}{63.63}{230}
\emline{120.12}{63.63}{231}{120.23}{65.45}{232}
\emline{120.23}{65.45}{233}{120.33}{68.67}{234}
\emline{117.33}{70.33}{235}{117.33}{67.33}{236}
\emline{120.33}{72.33}{237}{120.33}{66.67}{238}
\put(115.00,44.67){\makebox(0,0)[cc]{D3}}
\emline{119.00}{80.00}{239}{116.20}{79.16}{240}
\emline{116.20}{79.16}{241}{113.64}{78.31}{242}
\emline{113.64}{78.31}{243}{111.32}{77.48}{244}
\emline{111.32}{77.48}{245}{109.25}{76.64}{246}
\emline{109.25}{76.64}{247}{107.41}{75.81}{248}
\emline{107.41}{75.81}{249}{105.82}{74.98}{250}
\emline{105.82}{74.98}{251}{104.47}{74.16}{252}
\emline{104.47}{74.16}{253}{103.36}{73.34}{254}
\emline{103.36}{73.34}{255}{102.49}{72.52}{256}
\emline{102.49}{72.52}{257}{101.86}{71.70}{258}
\emline{101.86}{71.70}{259}{101.48}{70.89}{260}
\emline{101.48}{70.89}{261}{101.33}{70.08}{262}
\emline{101.33}{70.08}{263}{101.43}{69.28}{264}
\emline{101.43}{69.28}{265}{101.77}{68.48}{266}
\emline{101.77}{68.48}{267}{102.35}{67.68}{268}
\emline{102.35}{67.68}{269}{103.18}{66.89}{270}
\emline{103.18}{66.89}{271}{104.24}{66.09}{272}
\emline{104.24}{66.09}{273}{105.55}{65.31}{274}
\emline{105.55}{65.31}{275}{107.09}{64.52}{276}
\emline{107.09}{64.52}{277}{108.88}{63.74}{278}
\emline{108.88}{63.74}{279}{110.91}{62.96}{280}
\emline{110.91}{62.96}{281}{113.19}{62.19}{282}
\emline{113.19}{62.19}{283}{115.70}{61.42}{284}
\emline{115.70}{61.42}{285}{119.67}{60.33}{286}
\put(106.33,70.00){\makebox(0,0)[cc]{F1}}
\put(112.33,30.00){\makebox(0,0)[cc]{(b)}}

\put(93.33,80.00){\vector(-1,0){0.2}}
\emline{103.33}{80.00}{287}{93.33}{80.00}{288}
\put(113.33,80.00){\vector(1,0){0.2}}
\emline{103.00}{80.00}{289}{113.33}{80.00}{290}
\put(102.00,82.00){\makebox(0,0)[cc]{$v$}}
\emline{93.00}{50.00}{291}{93.00}{54.67}{292}
\emline{93.00}{57.33}{293}{93.00}{62.33}{294}
\emline{93.00}{65.33}{295}{93.00}{70.67}{296}
\emline{93.00}{73.67}{297}{93.00}{77.00}{298}
\emline{93.00}{78.67}{299}{93.00}{83.00}{300}
\put(92.67,46.50){\makebox(0,0)[cc]{$horizon$}}
\end{picture}

\vspace{-2.5cm}

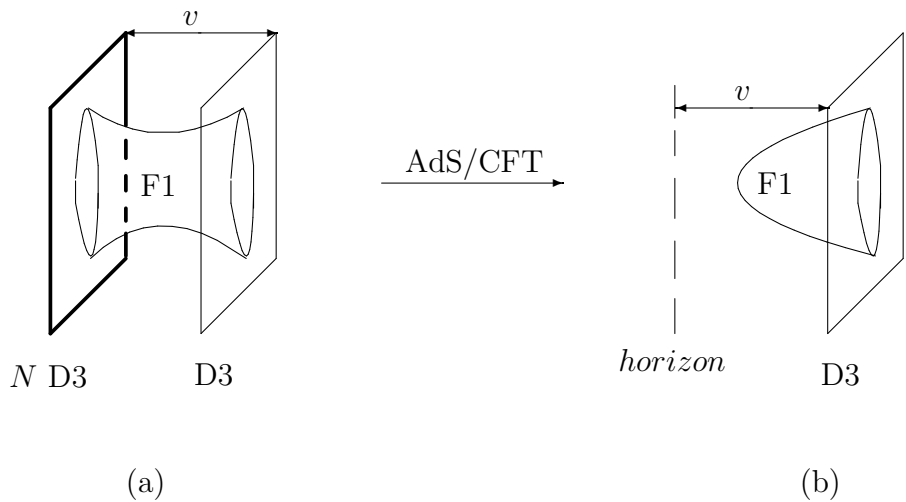
Fig.8: AdS/CFT correspondence at work: (a) Type IIB brane picture of the
W-boson pair creation induced by the U(1) part of the gauge strength
in \N4 theory with the U(N)$\ti$U(1) gauge group.
(b) Gravity dual picture: "cap"-like surface in the $AdS_5$ space.

\vspace{1cm}

The main idea here is as follows. We start with the bunch of
$N$ D3-branes in type IIB theory and flat background. The low energy
worldvolume theory is described by \N4 U(N) Yang-Mills theory. If we place
one additional D3-brane at the distance $v$ from the bunch,
this will result in U(N+1) gauge theory, spontaneously broken down to
U(N) $\ti$ U(1). Then, we can turn on constant electric field,
corresponding to the U(1) part of the gauge field and study creation
of the W-boson pair with the mass $v$ -- see Fig.8a.
In the large t'Hooft coupling constant limit we can also study this
process from the gravity dual point of view.
In order to do this we should replace a bunch of the D3-branes with
the $AdS_5$ space. $AdS_5$ curvature radius is given by

\be
R_{AdS}=(4\pi g_{YM}^2\a'^2N)^{1/4}
\ee
where $g_{YM}$ is Yang-Mills coupling constant. Instead of the "tube"-like
minimal surface in the flat space in Type IIB picture (Fig.8a) now we have
"cap"-like  minimal surface in curved space (Fig.8b).
All consequences of the F1 string interaction with the bunch of $N$ D3-branes
(including the strong coupling effects in associated gauge theory) are encoded
geometrically in the  Euclidean $AdS_5$ space metric:

\be
\label{metricz}
ds^2=R^2_{AdS}\frac{dr^2+r^2d\O^2_{3}+dz^2}{z^2}
\ee
where $(r,\O_{3})$ are the radial and angle coordinates on the D3-brane worldvolume,
while $z$ is a coordinate in the transverse (bulk) direction.
After the coordinate change in accordance with

\be
\label{r,t}
z=e^{\tau} \cosh^{-1}\r  \qquad
r=e^{\tau} \tanh\r
\ee
this metric takes the form:

\be
\label{metricrt}
ds^2=R^2_{AdS}(\sinh^2\r \ d\O^2_{3}+\cosh^2\r \ d\tau^2+d\r^2)
\ee
which will be useful in further calculations.

An important question concerns the  account of the electric field in the
gravity picture. One way is to include it directly into the metric
(see \cite{poleads}). Unfortunately, in this case it
is  hard to find explicit form of the minimal surface due to the lack of
the symmetry in the metric. Instead, we add boundary term (\ref{Sbndry})
to the effective action:

\be
\label{Sbndry2}
\oint d\sigma F_{01} X^0 \partial_{\sigma} X^1
\ee
Here $F_{\m\n}$ is the gauge strength. This term is nothing but an ordinary $B_{\m\n}$
term, which must be added to the Nambu-Goto action. It will be useful
to introduce "scalar" electric field $E$ with the help of the red-shift factor
from~(\ref{metricz}):

\be
E = F_{01} \frac{z^2}{R_{AdS}^2}
\ee
Then it is straightforward to see that the term (\ref{Sbndry2}) represents
the area \footnote{calculated in the $AdS_{5}$ metric (\ref{metricz})} of the disc
$\partial \Sigma$, which the minimal surface's boundary cut out
on the D3-brane. Therefore, we have the following effective  action:

\be
\label{STE}
S_{eff}=T_{F1} Area (\Sigma) + E Area (\partial \Sigma)
\ee
Of cource, is the surface $ \Sigma$ has two boundaries, there will be two boundary
terms in~(\ref{STE}).
To find the probability rate
we will minimize this action in the two steps. First, we will fix the boundary conditions and then find appropriate minimal surface. Second, we will extremize $S_{eff}$ with
respect to the boundary conditions.

Calculation of the probability rate resulting from the "cap"-like surface
(Fig.8b) is presented in section 4.2.
It is in fact similar to the circular Wilson loop calculation in \N4
SUSYM~\cite{esz,dg}, but in our case there are two essential differences.
First, there is a constant electric field on the D3-brane, which determines
certain angle  between F1-string worldsheet and D3-brane. Secondly,
we do not move this D3-brane up to infinity, since we need the mass of the W-boson
to be fixed.

Another interesting example concerns W-boson pair creation in the case when
U(N+2) gauge symmetry is spontaneously
broken down to the U(N)$\ti$U(2) gauge symmetry.
In the gravity dual picture we have two branes in the $AdS_5$ space and "tube"-like
string surface stretched between them. As we will see in section 4.3, in this case
our approach leads to some inconsistency in certain region of parameters.
This is presumably due to the fact that flat D3-brane, which we use as a probe,
is no more BPS-object when the electric field is turned on, and therefore some curving effects may occur.
Probably, it can be more instructive to study corresponding configuration
in the $AdS_3$ case.
This issues will be discussed elsewhere~\cite{GSS}.

\subsection{W-boson pair creation in \N4 SYM theory with the \\
U(N)$\ti$U(1) gauge group}

The area of the "cap"-like minimal surface with the circular boundary in the AdS space
without electric field was calculated in~\cite{Maldacena, Gross}.
The reader can find all necessary details in section 4.3.
Since there is  a point $r= \rho =0$ (a "pole") on a "cap"-like surface,
we must set $c=0$ in the integral of motion (\ref{intads}). Therefore,
equation for this surface in $(\rho,\tau)$ coordinates is just
$\tau= \const$. In terms of $(r,z)$ coordinates it has the form:

\be
r=a \tanh \r, \qquad z=a \cosh^{-1} \r, \quad \Rightarrow \quad
z^2 + r^2 = a^2
\ee
The role of the electric field is to determine the angle
between this surface and D3-brane. At zero value of the
field this angle is equal to $\pi/2$.
The action to be minimized is given by

\be
S=2\pi R^2_{AdS} (\sqrt{1+x^2}-1)-\pi R^2_{AdS} E x^2
\ee
where $x=\sinh \r$ defines the boundary radius of the surface.
The extremal value is

\be
x_*=\frac{\sqrt{1-E^2}}{E}
\ee
Therefore,

\be
S_{min}= \pi R^2_{AdS} \frac{(1-E)^2}{E}
\ee
and after restoring the $\a'$ and coupling constant dependence
the probability rate takes the form

\be
\label{U1rate}
w \propto \exp\left(- \sqrt{\pi N}
\frac{(1-2\pi \a' g_{YM}E)^2}{ E}\right)
\ee
Let us emphasize that it does not depend on the
W-boson mass or, in other words, on the position of the D3 brane in the AdS
space.

\subsection{W-boson pair creation in \N4 SYM theory with the \\
U(N)$\ti$U(2) gauge group }


As a first step we elaborate on the relation between the electric field and
the angle, at which the minimal surface touch D3-brane.
The effective action  in $(r,z)$ coordinates takes the form:

\be
\label{S(r,z)}
S=\pi R^2_{AdS} \left(2\int \limits_{z_R}^{z_L}\frac{r}{z^2}
\sqrt{1+(r')^2}dz-E\left(\frac{r_L^2}{z_L^2}+\frac{r_R^2}{z_R^2}\right) \right)
\ee
Here $r(z)$ is supposed to solve the equation of motion:

\be
\frac{d}{dz} \frac{rr'}{z^2 \sqrt{1+(r')^2}}=
\frac{ \sqrt{1+(r')^2}}{z^2}
\ee
with the appropriate boundary conditions: $r(z_{L,R})=r_{L,R}$. In order to
find the
minimum of (\ref{S(r,z)}) we should differentiate it with
respect to the $z_{L,R}$ and then set this derivatives  to zero.
By using the equation of motion we find:

\be
\label{Eband}
\frac{r'_L}{\sqrt{1+r'^2_L}}=E, \quad \frac{r'_R}{\sqrt{1+r'^2_R}}=-E
\ee
Therefore,

\be
\frac{dr}{dz}=\frac{E}{\sqrt{1-E^2}}=\xi
\ee
At the left (L) brane we have $\xi>0$ while at the right (R) brane
$\xi<0$. Note that we obtained the same junction (boundary) conditions as
in the weak coupling approximation (\ref{derivative}).
Moreover, we see that there exist the same critical value of the electric field as in
the 1+1 dimensional case. These  are the elementary tests on the consistency of our
approach.

As a next simple test we demonstrate how the Schwinger formula arises in the
framework of the AdS/CFT-correspondence. For this purpose we consider
production of the light W-bosons in the weak  electric
field\footnote{we assume that $\xi \ll v / R_{AdS} \ll 1$}.
The effective action has the form (see below):
\be
S=\pi R^2_{AdS} \left(2\int\limits_{x_L}^{x_R}
\frac{x^2 dx}{\sqrt{x^2(x^2+1)-c^2}}  -E (x_L^2+x_R^2) \right)
\ee
where

\be
x_{L,R}\approx c\sqrt{1+\xi^2}\mp\xi c^2
\ee
The value of the constant $c$ can be determined from the
condition that the distance $v$ between the left and right branes is
equal to

\be
\label{distance}
\frac{v}{R_{AdS}}=\2 \log \frac{1+x_R^2}{1+x_L^2}+
c\int\limits_{x_L}^{x_R}\frac{dx}{(x^2+1)\sqrt{x^2(x^2+1)-c^2}}
\ee
Suppose that the constant $c \gg 1$.
Then it is easy to show that the logarithm in (\ref{distance}) is
of order $\xi c$ while the integral is of order $\xi$ and therefore
irrelevant. Then,

\be
c \approx \frac{v}{2R_{AdS} \xi}
\ee
which implies that the field value and the W-boson mass should be small.
Now we can estimate the minimum of the  action
as follows:

\be
S_{min} \approx \frac{\pi v^2}{2 E}
\ee
This gives exactly the Schwinger dependence of the probability rate.

The rest of this section is devoted to the general case of the arbitrary
W-boson mass and electric field. More detailed and extended discussion
on the results given below will be presented elsewhere~(\cite{GSS}).
We take the metric in the form (\ref{metricrt}).
In this coordinates the horizon location is at  $\tau=\infty$.
The surface of rotation can be described in terms of the function $\r(\tau)$.
The area of the minimal surface of rotation with the fixed boundary conditions
in the $AdS_5$ space is given by the extremum of the functional:

\be
\label{Area}
A=2\pi R^2_{AdS} \int\sinh\r(\tau) \
\sqrt{\cosh^2\r(\tau)+\dot{\r}^2(\tau)} \ d\tau
\ee
It has the following integral of motion:

\be
\label{intads}
\frac{\sinh\r \ \cosh^2\r}{\sqrt{\cosh^2\r+\dot{\r}^2}}=c
\ee
where $c$ is a non-negative constant. Using this integral of motion,
we get the following equation for the minimal surface:

\be
\label{t(r)}
\int_{\tau_0}^{\tau_{\infty}} d\tau=\pm c\int
\limits_{\r_0}^{\infty}\frac{d\r}{\cosh\r \
\sqrt{\sinh^2\r \ \cosh^2\r-c^2}}
\ee
Here for simplicity we fix
one boundary condition at the horizon $(\tau_{\infty} \to \infty, \r =\infty)$ and
consider the area as a function of the other one: $\r(\tau_0)=\r_0$.
 The integral in
the rhs can be expressed via the elliptic integrals
(see, e.g. \cite{AS}):

\be
\label{tau-ell}
\tau_{\infty}-\tau_0=\pm \frac{c}{a}\left(F(a/x_0,ik) -
\Pi(a/x_0,-1/a^2,ik )\right)
\ee
where $x_0=\sinh\r_0$ and

\be
a=\sqrt{\frac{\sqrt{1+4c^2}-1}{2}}, \qquad
k=\sqrt{\frac{\sqrt{1+4c^2}+1}{\sqrt{1+4c^2}-1}}
\ee
Elliptic integrals in (\ref{tau-ell}) are defined as

\be
F(x,k)=\int\limits_{0}^{x} \frac{dt}{\sqrt{(1-t^2)(1-k^2 t^2)}}
\ee
and

\be
\Pi(x,n,k)=\int\limits_{0}^{x} \frac{dt}{(1-nt^2)\sqrt{(1-t^2)(1-k^2
t^2)}}
\ee
We can plug (\ref{t(r)}) into (\ref{Area}) and obtain

\be
A_{min}[x_0,c]=2\pi R^2_{AdS} \int\limits_{x_0}^{x_{\infty}}
\frac{x^2d x}
{\sqrt{x^2(1+x^2)-c^2}}, \quad x=\sinh\r
\ee
It  can be also expressed via the elliptic integrals:

\be
A_{min}[x_0,c]=2\pi
R^2_{AdS}\left(x_{\infty}-\sqrt{1+x_0^2-c^2/x_0^2}+aF(a/x_0,ik)-
aE(a/x_0,ik) \right)
\ee
where

\be
E(x,k)=\int\limits_{0}^{x} \frac{\sqrt{1-k^2t^2}}{\sqrt{1- t^2}} dt
\ee
As we discussed before, in order to
turn on constant $U(1)$ part  $E$ of the  gauge field on the probe
brane, we should include the  term

\be
S_E=- 2\pi E \int  \frac{rdr}{z^2_0}=
-E R_{AdS} \sinh^2 \r_0
\ee
Therefore, we should minimize the action

\be
S=T_{1,0}A_{min}-\pi R^2_{AdS} E x_0^2
\ee
with respect to the $x_0$  assuming that the distance $v$
between the boundaries of the surface is a constant.
The latter is given by

\be
\label{v}
v=R_{AdS}\int_{z_0}^{z_{\infty}}\frac{dz}{z}=
R_{AdS}\left( \2 \log\frac{1+x_{0}^2}{1+x_{\infty}^2} \pm
\frac{c}{a}\left(  F(a/x_0,ik) -
\Pi(a/x_0,-1/a^2,ik )\right)\right)
\ee
Then, it is straightforward to rewrite  $dr/dz$ derivative
in $(\r, \tau)$ coordinates using (\ref{r,t},\ref{t(r)}):

\be
\label{xi}
\frac{dr}{dz}=  \frac{cx \pm \sqrt{x^2(x^2+1)-c^2}}
{c\mp x \sqrt{x^2(x^2+1)-c^2}}
\ee
The equation (\ref{Eband}): $dr/dz=\xi$ generally has four solutions:

\be
\label{solution}
x^{(\pm,\pm)}=\frac{-1\pm \sqrt{1\pm 4 \xi c \sqrt{1+\xi^2}}}{2\xi}
\ee
The sign which we should take in (\ref{xi}) is the same as
in (\ref{t(r)}) and is equal to

\be
\pm = \frac{\sgn(1+\xi x)}{\sgn(\xi-x)}
\ee
From the analysis of these branches one can see that
if one starts with some value $E$ of the feld and distance $v$
and then begins to increase them, the regime changes. At certain values of the parameters
the solution disappears (becomes complex). However, the effective action
do not vanishes at this point. This indicates that our approach fails
at certain region of the parameters.















\section{Pair production in the  noncommutative
gauge theory and NCOS theory}

\subsection{On the pair production in NCYM theory}

In this section we examine the Schwinger type processes
in the noncommutative
theories which can be considered as the certain
limits of the theories on D-branes with the background
magnetic \cite{sw} or electric \cite{sst,gmms}
fields (see \cite{dn} for a review). Actually the theories
with
electric or magnetic backgrounds can be treated as dual to
each other and the discussion
concerning the relevant duality group
can be found in \cite{russo,verlinde}. We would like  to consider
nonperturbative aspects of the NCYM and
NCOS theories. The motivation can be explained
recalling the field theory counterpart. It is well known
that the electrodynamics in the constant electric
field is unstable theory with respect to the
nonperturbative pair production. The Schwinger
mechanism screens the electric field and
finally the stable vacuum state has
vanishing electric field. Therefore it is natural
to ask if NCYM and NCOS are nonperturbatively
stable theories. We will discuss the possible modes
of instability in both theories.

Consider first the Dp branes
with the magnetic backgrounds which can be obtained
from the space $B_{\mu \nu}$ field in the bulk after
the appropriate gauge fixing.
Another way to represent magnetic field
is to assume that there is the nonvanishing density
of D1 branes on Dp brane worldvolume.
The  limit leading  to the NCYM theory
\cite{sw}  can be described as follows. Suppose that we start
with the flat closed string metric $g_{ij}$  and the weak
closed string coupling $g_s$. Then the parameters
of the open string  theory are
\be
G_{ij}=g_{ij} - (2\pi \alpha^{'})^2 (Bg^{-1}B)_{ij}
\ee
\be
\theta^{ij}=  - (2\pi \alpha^{'})^2 \left(\frac{1}{g + 2\pi \alpha^{'}B} B
\frac{1}{g - 2\pi \alpha^{'}B}\right)^{ij}
\ee

\be
G_s=g_s  \left(\frac{\det G_{ij}}{\det(g_{ij} + 2\pi \alpha^{'}B_{ij})}\right)^{1/2}
\ee

To get the Seiberg-Witten limit leading to the NCYM theory one has
to take $ \alpha^{'} \rightarrow 0$ with $G_{ij}, \theta^{ij} , G_s$
kept fixed \cite{sw}.
To this aim in the flat space one has either to vanish closed string metric
or to take $B \rightarrow \infty$. The latter limit is more natural
physically however we would like to discuss its reliability.
It was assumed in \cite{sw} that the limit $B \rightarrow \infty$
can be performed safely since there is no critical magnetic field.
We would like to argue that it is not the case. We have already
meet  the critical magnetic field in Section 2 where the
nonperturbative production of the monopole pair was discussed.
Even more simple argument involves the consideration of the
D1 string with the monopole pair at its ends in
the magnetic field. The critical
magnetic field corresponds to the situation when
the force pulling apart the monopoles at the ends exactly
balances the tension of the D1 string. The corresponding
critical magnetic field coincides with that one
amounted from the Schwinger process\footnote{
The existence of the critical magnetic field
has been noted before in~\cite{Rey} in another context.}.

Let us show that  SW limit encounters the
problem indeed. To this aim let us take
the following  magnetic field
and closed string metric in $(x_2,x_3)$ directions
\be
g_{ij}= g_{cs}\diag(1,1),\qquad B_{ij}=B\epsilon_{ij}
\ee
The open string parameters are

\be
G_{ij}= G \diag(1,1) \qquad
\theta^{ij}= \theta \epsilon_{ij}
\ee
where
\be
G=(1+ \tilde{B}^2) \qquad
\theta=\frac{\tilde{B}}{B_{cr}(1+\tilde{B}^2)} \qquad
\tilde{B}=\frac{B}{B_{cr}}.
\ee
From the relation
\be
\label{critB}
2\pi \tilde{B} \a' = G\theta
\ee
which does not involve  small parameter $g_{cs}$ at all
it is clear that if there exists the critical magnetic field
the naive decoupling of the stringy degrees of freedom is impossible.
Let us emphasize that the critical magnetic field value
does not go to the infinity in the $\a' \rightarrow 0$ limit
due to the proportionality to  $g_{cs}$.
Note that the open string coupling at the critical
magnetic field equals to  $g_s$.

Therefore we can investigate
the pair production in pure noncommutative gauge theory
taking the SW limit of the
nonperturbative amplitude in the theory
on the D3 brane in the magnetic background.
Let us discuss the possible unstable modes in U(1) NCYM
theory.  In this theory there are magnetically charged
U(1) noncommutative monopoles \cite{gn} which however have
infinite masses and correspond to semiinfinite D1 strings
ended
on D3 brane  as well as open finite D1 strings. The probability
of the monopole production vanishes however it seems  that the
probability of the pair of magnetically
charged D1 production survives similarly the Bachas-Porrati
description of the annulus partition function. Moreover
we could also expect that similarly to the electric case
\cite{burgess} there could be a sort of classical
instability for the magnetically neutral D1
open string at the critical magnetic field.

The static  monopole
in U(2) NCYM theory is represented by the tilted D1 string
stretched between two D3 branes. The angle between the
D1 and D3 branes is determined by the tension condition.
Unlike the commutative case the monopole state looks
as a magnetic dipole whose size is determined by
the noncommutativity \cite{gn}.
Once again we are looking for the
Euclidean solution corresponding to the monopole
pair production. In this case we are producing
the pair of dipoles. It seems that there
is no Euclidean configuration corresponding
to the production of U(2) noncommutative monopoles
if the magnetic fields on the two D3 branes  are different:
$B_1\neq B_2$.

\subsection{On the nonperturbative processes in NCOS theory}
Let us turn to the
S-dual of noncommutative SYM theory, namely
NCOS theory with
nearly critical electric field
and remaining stringy degrees of freedom \cite{gmms,sst}.
Now take the electric field
and closed string metric in $(x_0,x_1)$ directions
\be
g_{ij}= g_{cs}\diag(-1,1),\qquad B_{ij}=E\epsilon_{ij}
\ee
The open string parameters are
\be
G_{ij}= G \diag(-1,1) \qquad
\theta^{ij}= \theta \epsilon^{ij}
\ee
where
\be
G=(1- \tilde{E}^2) \qquad
\theta=\frac{\tilde{E}}{E_{cr}(1-\tilde{E}^2)} \qquad
\tilde{E}=\frac{E}{E_{cr}}.
\ee
The relation between the parameters looks similarly to the
magnetic case
\be
\label{critE}
2\pi \tilde{E} \a' = G\theta
\ee
and the theory near the critical electric field defines the NCOS theory.

The essential difference with the magnetic case
is that in the finite $g_s$ case the effective
coupling constant of the NCOS theory vanishes.
However one could consider
the following limit
\be
g_s \rightarrow \infty, \qquad  g_s^2 \alpha^{'} \ {\rm fixed}
\ee
The parameters of the NCOS theory are related
to the parameters of IIB theory as follows
\be
 g_s^2 \alpha^{'} = G_s^4  \alpha^{'}_{eff}
\ee
\be
\alpha^{'} \tr F = 1 - \frac{\alpha^{'}}{2\alpha^{'}_{eff}}
\ee
where $F$ is the  field strength and $G_s$ is the coupling
of the NCOS theory. In the limit discussed above
the field approaches the critical value $\alpha^{'}\tr F = 1$.

Consider the possible instability modes. There
is the evident classical instability in the
electrically neutral sector in U(1) theory.
Moreover it is clear
that in U(1) theory there is almost unsuppressed
Bachas-Porrati probability
of the creation
which for each stringy mode involves  in this limit the factor
\be
w\propto \exp\left(\frac{\const}{\log(E-E_{cr})}\right)
\ee
In the nonabelian case, for instance with the U(2) gauge group, one could ask about
the production of the pair of "W-bosons", that is  strings stretched
between two D3  branes in d=4 theory.  From the consideration in Section 2
it seems that the probability doesn't vanish if
the fields on  each U(1) factors are different $E_1\neq E_2$.

The behaviour of the NCOS theories at the
nonzero temperature was recently considered in \cite{gukov,barbon}.
It was shown that the
Hagedorn transition corresponds to the
temperature when getting off  the neutral  strings from the
NCOS worldvolume becomes favorable. The quantum
instability considered here doesn't influence the
Hagedorn transition for the neutral sector but
would affect the transition in the charged sector.

Finally, note that one could add fundamental matter
represented, for instance, by the D7 branes. To discuss the
Schwinger pair production of the fundamental matter we can use
the behaviour of the fundamentals found above. Near the
critical electric field the Euclidean configuration
responsible for this process looks like an almost
noncompact surface approaching D3 brane and having the cusp
at the attachment point to D7 brane.

\section{Conclusion}

In this paper we investigated the Schwinger
type amplitudes in the context of the
theories on the D-branes  worldvolumes.
It was shown that in the weak coupling
approximation the picture of the particle
production is very transparent and
the agreement with the string production
in the external field was found.
We mainly focused in this paper on the
theories with the maximal SUSY which
corresponds to the field theory on the
parallel D branes. However, we expect that
the same approach works also for the theories
with the less amount of SUSY where
the curved geometry of the brane worldvolumes
has to be taken into account.
A new
manifestation of the critical electric
and magnetic fields was found. The calculation
of the stringy deformed monopole pair production
can provide some guess for the derivation
of the partition function of the D-string.
Some comments concerning the behaviour
of the theories at the finite temperature
were presented however the complete
analysis will be considered elsewhere.

One can consider more general
examples of the Schwinger type processes which
can involve the branes of different dimensions.
For instance, the important
generalization of these processes with the additional
particles (branes) in the final states
has been recently used for the description
of the Brane World spontaneous production~\cite{gs2}
in the higher rank field.
It would be interesting to generalize
such processes for the case when the Brane World created
along this way carries the additional
particles or strings.

We elaborated the pair production
processes in the context of
AdS/CFT correspondence and found
the probability of the process at the
strong coupling limit. Surprisingly enough
the critical electric field also manifests itself in
this calculations. We used the simplified version
of AdS/CFT correspondence
when the electric field is introduced via a
term in the effective action. This works well in $AdS_3$ case
however the exact modification of the metric
by the external field has to be taken into account in the
higher dimensions. We plan to discuss this point
elsewhere.

The noncommutative theories are the
natural playground for the application of
Schwinger processes  hence we
investigated the issue of the nonperturbative
stability of the NCYM and NCOS theories. It appears
that in both cases the potential modes of instability
are indicated. Contrary to the widely accepted
viewpoint concerning the absence  of the critical
magnetic we claim that it does exist and has
very clear interpretation. We have mentioned
that such critical magnetic field can
be the obstacle for the Seiberg-Witten limit
to exist. Furthermore, it is possible that Schwinger type processes
give nonperturvatibe (nonanalytical in the field) corrections
to the Born-Infeld effective action
describing the brane worldvolume theory.
However, to make more definite
conclusion concerning this point
additional analysis is required.

One more interesting outcome of this article
is the  discovery of a few novel
nonperturbative processes of the BPS
particles  decay in the external fields. We can not
exclude that these amplitudes can have some
phenomenological implications in the early Universe.
For instance the strong magnetic fields
could produce the magnetically
charged particles from the electrically
charged ones with the substantional rate.

We would like to thank S.Gukov, I. Klebanov and N.Nekrasov
for the useful discussions. The work of A.S.G. was supported by
grant INTAS-00-00334 and  CRDF-RP2-2247 ,
the work of K.A.S. was supported in part by the Russian
President's grant 00-15-99296, RFBR grant 98-02-16575 and INTAS grant
99-590, the work of K.G.S. was supported by grant INTAS-00-00334
and CRDF- RP1-2108.

\end{document}